\documentstyle[galley,graphicx,epsf]{mn}

\begin{document}

\title {The Chemical Compositions of Galactic Disk F and G Dwarfs}

\author[Reddy et al.] 
        {Bacham E. Reddy, Jocelyn Tomkin, David L. Lambert, Carlos Allende Prieto \\
        Department of Astronomy, University of Texas, Austin, Texas 78712}


\pagerange{\pageref{firstpage}--\pageref{lastpage}}

\maketitle

\label{firstpage}

\begin{abstract}
Photospheric abundances are presented for 27 elements from carbon to europium in
181 F-G dwarfs from a differential LTE  
analysis of high-resolution and high signal-to-noise spectra.
Stellar effective temperatures ($T_{\rm eff}$) were adopted from 
an infrared flux
method calibration of Str\"{o}mgren photometry. Stellar surface
gravities ($g$) were calculated from {\it Hipparcos} parallaxes and
stellar evolutionary tracks. Adopted $T_{\rm eff}$ and $g$ values
are in good agreement with spectroscopic estimates.
Stellar ages were determined from  evolutionary
tracks.
Stellar space motions ($U, V, W$) and a Galactic potential were used to
estimate Galactic orbital parameters. These show that the vast
majority of the stars belong to the Galactic thin disc.

Relative abundances expressed as [X/Fe] generally confirm previously
published results. We give results for C, N, O, Na, Mg, Al, Si, S,
K, Ca, Sc, Ti, V, Cr, Mn, Co, Ni, Cu, Zn, Sr, Y, Zr, Ba, Ce,
Nd, and Eu.  The $\alpha$-elements -- O, Mg, Si, Ca, and
Ti -- show [$\alpha$/Fe] to increase slightly with decreasing [Fe/H].
Heavy elements with dominant contributions at solar metallicity from
the $s$-process show [$s$/Fe] to decrease slightly with
decreasing [Fe/H]. Scatter in [X/Fe] at a fixed [Fe/H] is entirely
attributable to the small  measurement errors, after excluding the
few thick disc stars and the $s$-process enriched CH subgiants.
Tight limits are set on `cosmic' scatter. 
If a weak trend with
[Fe/H] is taken into account, the composition of a thin disc star
expressed as [X/Fe] is independent of the star's age and birthplace
for elements contributed in different proportions by
massive stars (Type II SN), exploding white dwarfs (Type Ia SN),
and asymptotic red giant branch stars.

By combining our sample with various published studies, comparisons
between thin and thick disc stars are made. In this composite sample,
thick disc stars are 
primarily identified 
by their $V_{LSR}$ in the range $- 40$ to $-100$ km s$^{-1}$. These are very old
stars with origins in the inner Galaxy and metallicities [Fe/H] $\leq -0.4$.
At the same [Fe/H], the sampled thin disc stars have
$V_{LSR} \sim 0$ km s$^{-1}$, and are generally younger with a birthplace
at about the Sun's Galactocentric distance. 
In the range
$-$0.35 $\geq$ [Fe/H] $\geq$ $-$0.70, well represented by present thin and thick 
disc samples, [X/Fe] of the thick disc stars is greater than that of thin
disc stars for Mg, Al, Si, Ca,  Ti, and Eu. [X/Fe] is very
similar for the thin and thick disc for -- notably -- Na, and
iron-group elements. Barium ([Ba/Fe]) may be underabundant
in thick relative to thin disc stars.  These results extend
previous ideas about composition differences between the thin and
thick disc.

\end{abstract}


\section{Introduction}
Lower main sequence stars have lifetimes comparable to the
age of the Galaxy and presumably 
atmospheric compositions that are essentially identical
to those of their natal interstellar clouds. Spectroscopic and photometric
analysis of these stars provides a sensitive
probe of the major processes that have shaped the chemical evolution of our
Galaxy.
This paper, which describes a survey of 181 F and G
main sequence stars,
was inspired by Edvardsson
et al.'s (1993, hereafter EAGLNT) analysis of abundances
for 13 elements for 189 F and G disc
dwarfs with metallicities in the range $-$1.1 $\leq$ [Fe/H] $\leq$ +0.25.
We sought to examine more closely several conclusions
advanced by EAGLNT.

One conclusion concerned the variation of chemical composition with
the distance of a star's   birthplace from the Galactic centre.
EAGLNT gave estimates of this distance ($R_{\rm m}$) derived from a star's
kinematics and a model of the Galactic potential. A striking dependence
on $R_{\rm m}$ was found for the relative abundances of $\alpha$-elements (Si
and Ca) and iron. As was already known (cf. Lambert 1989; Wheeler, Sneden,
\& Truran 1989; McWilliam 1997), [$\alpha$/Fe] increases with decreasing
[Fe/H], rising from [$\alpha$/Fe] = 0 at [Fe/H] = 0 to about
0.3 at [Fe/H] = $-$1. EAGLNT found that the trend of [$\alpha$/Fe] with [Fe/H]
depends on $R_{\rm m}$, being more marked for small $R_{\rm m}$ than large $R_{\rm m}$. 
They
interpreted this dependence of [$\alpha$/Fe] on $R_{\rm m}$  
at a given [Fe/H] as due to an early rapid rate of star formation
in the inner parts of the Galactic disk resulting
in Type II supernovae dominating the enrichment of the
interstellar gas to a greater extent than they did locally where
the more slowly evolving Type Ia supernovae have been important
contributors.
Models of the Galactic chemical evolution have predicted
abundance gradients of the kind inferred by Edvardsson et al.
(e.g., Chiappani, Matteucci, \& Gratton 1997).

Others have examined this and other of EAGLNT's conclusions. Fuhrmann (1998)
finds that 
[Mg/Fe] gets successively smaller in halo, thick disc, 
and  thin disc stars, and there
is a segregation of [Mg/Fe] between two disc populations such that even at
the same [Fe/H] their [Mg/Fe] are distinct.  Chen et al. (2000), who
analysed a sample of 90
F and G disc dwarfs for 13 elements, found a group of stars in the
metallicity range $-$1.0 $\leq$ [Fe/H] $\leq$ $-$0.6 having small $R_{\rm m}$ ($\leq$
7 kpc) that are older than other disc stars and probably belong to the
thick disc.

Our present survey covers 181 nearby F and G dwarfs observed with
the McDonald Observatory's 2.7-m telescope and 2dcoud\'{e}
spectrometer at a resolving power of about 60,000 with broad spectral
coverage, and at S/N ratios of 300 to 400 for most stars. The wavelength
coverage and S/N ratios are a significant improvement over EAGLNT's
observations, which covered four or five spectral regions of 100~\AA\
each at a S/N ratio of about 200. Also, EAGLNT's  northern stars were
observed at a resolving power of about 30,000, and only the southern stars
were observed at the resolving power of 60,000.
Our new spectra lead to more accurate abundances
for more elements. Our analysis benefits also from the
use of improved fundamental parameters for the stars. In particular,
the effective temperatures are determined from the $b - y$ colour and
a recent calibration based on the infrared flux method, while
surface gravities come from a comparison of the stars' positions in the
colour-magnitude diagram, which are precisely fixed by the {\it Hipparcos}
parallaxes, with theoretical isochrones.

\section{Observations}

\subsection{Stellar Spectra}

Programme stars were selected from the $uvby\beta$ catalogue of Olsen (1983,
1988).  Among the selection criteria were  an effective temperature in the
range 5600 K to 7000 K, and a surface gravity indicative of little or moderate
evolution off the zero age mainsequence (ZAMS) (see Fig~1.).

All the observations were made at the Harlan J. Smith 2.7-m telescope at
McDonald Observatory, using 
the 2dcoud\'{e} echelle spectrometer (Tull et al. 1995)
with a 2048 $\times$ 2048 pixel Tektronix CCD as detector. A resolving
power of $\approx$ 60,000 was attained. Spectral coverage was 
complete from 4000 to 5600 \AA\ and
substantial but incomplete from 5600 \AA\ to about 9000~\AA.
In order to minimize the influence of  cosmic rays, two observations
in succession, rather than one longer observation, were generally
made of each star. From about 5500~\AA\ to about 9000~\AA\ the extracted
stellar spectra have a typical S/N ratio of about 400, while at
wavelengths shorter than about 5500\AA\ the S/N ratio
decreases with decreasing wavelength. 
We also observed the
asteroid Iris in order to have a solar spectrum recorded under similar
circumstances as the stellar spectra.
The data were processed and wavelength calibrated in a conventional
manner with the IRAF\footnote{IRAF is distributed by the National
Optical Astronomical Observatories, which is operated by the
Association for Universities for Research in Astronomy, Inc., under
contract to the National Science Foundation.} reduction package.
Double-lined spectroscopic binaries and broad-lined stars ($v\sin i \geq
20$ km s$^{-1}$) were  dropped from the programme. The remaining 181 stars
were subjected to an  abundance analysis, and are listed in Table\,1.

Absorption lines suitable for measurement were chosen for having 
clean line profiles,
as judged by inspection of the solar flux spectrum at 
extremely high resolving power and
S/N ratio (Kurucz et al. 1984), that could be reliably
measured in all, or most of, the programme stars. Moore, Minnaert, \&
Houtgast (1966) was our primary source for line identification.
The equivalent width of each line was measured with the IRAF $splot$ measurement
option most suited to the situation of the line. This was usually the
fitting of a single (or multiple) Gaussian profiles to the stellar line,
but for stronger lines with significant damping wings a Voigt
profile was used; for a few lines direct integration provided the
best method of measurement and was preferred. Table\,2 gives basic information
for the selected lines; the list includes 170 lines of 27 elements.
The spectrum of Iris, which was reduced and measured in the same manner as the
programme stars, provided solar equivalent widths.

\subsection{Stellar Kinematics}

The space motions of the programme stars can be used to calculate
their Galactic orbits.  In order to determine the space motions
we need the stellar distances, proper motions, and radial
velocities.

Parallaxes and proper motions for nearly all the programme stars
are available from the {\it Hipparcos} Catalogue (ESA 1997).
All the stars in our sample are within 150 pc from the Sun,
so their {\it Hipparcos}
parallaxes are accurate; the average percentage error 
is 6.1 $\pm$ 3.2.
The proper motion errors are much less.  For the few (7)
programme stars not in the Hipparcos catalogue $uvby\beta$
photometry was used to calculate photometric distances
following the prescriptions of EAGLNT.  A comparison of
photometric and Hipparcos distances showed that
the average ratio d(phot)/d(Hip) is
1.10 $\pm$ 0.15 (20 stars, standard deviation).

Accurate CORAVEL radial velocities have been taken from the
survey of kinematical data for F-G dwarfs in the solar
neighborhood by Pont et al. (1999).  These data were
kindly provided by J. Andersen prior to publication.
The CORAVEL data show that 19 of the programme
stars\footnote{These stars are: HD 3454, 6840, 22521, 85902,
89010, 101472, 112756, 124819, 156635, 192145, 200580, 201444,
201639, 204559, 210718, 210985, 219497, 220908, and 225239.}
have variable radial velocities, presumably a result of their
membership in low-amplitude spectroscopic binary systems. 
Small corrections ($< 1$ km s$^{-1}$) to the observed velocities due
to gravitational or convective shifts were neglected.

The space velocities with respect to the Sun were then calculated.
A solar motion (-10.0,+5.2,+7.2) km s$^{-1}$ in
(U,V,W) \footnote{In this study U is defined to be positive in
the direction of the Galactic anticentre.}, derived from
{\it Hipparcos} data by Dehnen \& Binney (1998), was adopted in
adjusting the space velocities to be with respect to the
Local Standard of Rest.  
Table~1 gives
$(U_{\rm LSR}, V_{\rm LSR}, W_{\rm LSR})$.
Our space velocities may be
compared with those of Chen et al. (2000) who use the same
solar motion. For the 19 stars
in common to our investigation and theirs, which are also in
the {\it Hipparcos} catalogue and also show no sign of 
radial-velocity variation so their space velocities are based on
identical proper motions and distances and very similar
radial velocities, the (U,V,W) velocities agree to about 1 km s$^{-1}$,
or better.

The Galactic orbital parameters, $R_{\rm p}$ and $R_{\rm a}$ (peri- and apogalactic
distances), Z$_{max}$ (maximum distance from the Galactic plane), and the $e$ (orbital eccentricity)
were computed using a Galactic potential integrator developed 
by D. Lin (1999, provided by Jon P. Fulbright with the kind 
permission of D. Lin).
 	
In computing orbital parameters we have adopted the above solar space motion,
a solar galactocentric distance of 8.5 kpc, and
a solar circular velocity of 220 km s$^{-1}$. 
Grenon (1987) argues that $R_m = (R_p + R_a)/2$ is a
likely stable quantity and, hence, a measure of a star's
birthplace. 
In Table\,1, we give the orbital parameters
$R_{\rm m}$, $Z_{\rm max}$, and $e$.

We have only 3 stars
in this study which are common with EAGLNT's sample. The orbital
parameters  derived in this study are different from EAGLNT's 
by $\leq$ 5$\%$, except
Z$_{max}$ which differs by $\sim$ 25$\%$. The differences
become  smaller ($\leq$ 2$\%$) if we adopt the same values for the 
Sun as that of EAGLNT [EAGLNT used (U,V,W) = (-10, +6, +6 km s$^{-1}$), 
8.0 kpc for
galactocentric distance, and 226 km s$^{-1}$ for the solar circular velocity].

\section{Analysis}

\subsection{Introduction}

Elemental abundances are derived from an LTE  analysis of
equivalent widths using the code MOOG (Sneden 1973). ATLAS9
(Kurucz 1998) plane-parallel, line-blanketed, flux-constant LTE
model atmospheres with convective overshooting are used.
The models are linearly interpolated for the appropriate
values of the fundamental
atmospheric parameters ($T_{\rm eff}$, $\log g$, 
[M/H]\footnote{The usual bracket notation is used throughout the paper for
the abundance of an element M with respect to hydrogen 
[M/H]= $\log \frac {(N_{\rm M}/N_{\rm H})_{\star}} {(N_{\rm M}/N_{\rm H})_{\odot}} $ 
where, N represents number density.}, 
$\xi_t$), which 
are determined independently  of the spectroscopy.
The effective temperatures $T_{\rm eff}$ and metallicities [M/H]
are derived from $uvby\beta$-photometry. 
The surface gravities $g$ are determined from the comparison of the position
of the star in the $(B-V) - M_{\rm V}$ plane with  calculations of stellar evolution,
using {\it Hipparcos} parallaxes to determine $M_{\rm V}$. 
The microturbulence $\xi_t$ is set by an empirical
relation between $\xi_t$, $T_{\rm eff}$, and $\log g$ derived from
spectroscopic analysis of a subset of the programme stars. The large number
of lines available for elements such as iron allows an independent
spectroscopic determination of $T_{\rm eff}$ and $\log g$, and, hence,
a comparison of the photometric estimates of these quantities.

\subsection{Selection of  The Model Atmosphere Grid}

 Over the last decade we have witnessed exciting developments
 in stellar atmosphere modeling well 
 beyond classical LTE one-dimensional models. 
 Full NLTE structures (e.g. Hauschildt et al. 1999), LTE 3D time-dependent 
 hydrodynamical models (e.g. Asplund et al. 2000), and 1.5D and 3D  
 NLTE radiative transfer calculations (e.g.  Shchukina \& Trujillo Bueno  2001) 
 are examples of the recent advances. Application of these 
 modeling techniques
 to a large sample of stars, and elements is still unpractical for several
reasons: the new
 models are available only for a few values of the atmospheric
 parameters; NLTE calculations with realistic model atoms are generally
 time consuming and, for some species, unreliable, due to uncertainties in 
 the atomic data. Recent studies (Nissen et al. 2002; Chen et al. 2002)
 provide comparisons for some elements of abundances derived from 1D and 3D models.
 For stars with [Fe/H], $T_{\rm eff}$, and log $g$ corresponding to our
  dwarfs, these authors find
 [O/Fe], [S/Fe], and [Si/Fe]
 from 3D are lower by less than 0.04~dex than results from an equivalent 
 1D model. The differential effect on such abundance
 ratios across our sample of similar stars should be very small. Effects of replacing
 1D models by 3D models  may vary from element to element.
 While we are confident that the 
 the situation will change in the near future, we have considered only
well-tested flux-constant 1D model atmospheres: 
the MARCS models originally developed by Gustafsson et al. (1975), and
the ATLAS9 models incorporating improvements to the treatment of
convection and convective overshoot (Kurucz 1998).
We also tested recently developed atmospheric models known as 
NEXTGEN models by Hauschildt, Allard, \& Baron (1999).

In comparisons against observations of the Sun, the ATLAS9 solar
model fares better than the MARCS solar model. This is clearly the
case for limb darkening in the continuum at optical and near-infrared
wavelengths (Blackwell et al. 1995). A better fit to limb darkening data
is achieved with the empirical model atmosphere known as the
Holweger-M\"{u}ller (1967,1974) model; a not surprising result given that
the model was derived in large part from limb darkening measurements.
Models may also be compared by their ability to fit those Fe\,{\sc i} lines
having an accurately determined $gf$-value from laboratory experiments.
Iron abundances derived from lines of different lower excitation
potential are less dependent on the lower potential in the case of the
ATLAS9 than for the MARCS model (Blackwell et al. 1995). The empirical
Holweger-M\"{u}ller model, as revised by Grevesse \& Sauval (1999),
and the MISS (Allende Prieto et al. 2001a) empirical
model,
also return iron abundances that are independent of the lower
excitation potential. Ionization equilibrium, as measured by the
iron abundance derived from Fe\,{\sc i} and Fe\,{\sc ii} lines, is
well satisfied by the ATLAS9, and the Grevesse-Sauval 
or the MISS  empirical models, but less well by the MARCS and NEXTGEN models.
ATLAS9 models come in two flavors - with and without convective
overshooting considered as part of the treatment of convection.
Castelli et al. (1997) suggest that the model with overshooting
(OVER model) reproduces observed properties (limb darkening etc.)
of the Sun better than the model without overshooting (NOVER model).

Our choice
of OVER ATLAS9 over MARCS models for the abundance analysis is based on 
the solar comparison.
It should be noted, however, that almost all of the programme
stars are within $\pm$500~K of the Sun's temperature. 
If the difference between
an ATLAS9 and a MARCS model is partly attributable to the different treatments
of line blanketing, the differences in abundances derived from them should be smaller for the
typical  star in our sample than they are for the Sun.
Similarly, the small differences between NOVER and OVER ATLAS9
solar models will become even smaller for the programme stars
where we are concerned with differential abundances relative to Sun.
We comment below on the
effects of replacing the preferred OVER with the NOVER models.

\subsection{Fundamental Atmospheric and other Parameters}

Three of the four parameters listed above are used to select
a model from the OVER ATLAS9 grid. These -- $T_{\rm eff}$, $\log g$,
and [M/H]--  are first determined from photometry,  and checked
subsequently against the spectroscopic analysis. Microturbulence, the
fourth parameter, is only determined 
spectroscopically. In addition, it is not used in the selection
of a model from the grid; the grid was computed for single
value of the microturbulence ($\xi_t$ = 2.0 km s$^{-1}$) which
is fairly representative of values determined here spectroscopically.

\subsubsection{The Effective Temperature}

The $uvby\beta$ photometry, especially the $b - y$ colour, is used to
determine the $T_{\rm eff}$. First, we must consider and, if necessary,
correct for the effects of interstellar reddening. The programme stars
are all within 150 pc of the Sun, with the majority  within
100 pc. Interstellar reddening is negligible within 100 pc (Schuster
\& Nissen 1989). In order to check for reddening at greater distances
we considered the average value and distribution of  $E{\rm (b - y)}$ for
stars closer than 100 pc and likewise for the stars further away than
100 pc.  $E{\rm (b - y)}$ came from the observed $b - y$ and the
unreddened $(b - y)_0$ derived from the $\beta$ index, which is
unaffected by reddening, together with Olsen's (1988) calibration.

For stars within 100 pc, the average $E{\rm (b - y)}$ is very small:
$E{\rm (b - y)} = -0.005 \pm 0.010~mag$ from 170 stars, and the distribution
is essentially Gaussian with the dispersion explained by the errors in the
photometry. For the more distant 21  stars, $E{\rm (b - y)}$ is positive:
$E{\rm (b - y)} = +0.010 \pm 0.013~mag$, and the distribution is asymmetric  with
a tail of positive $E{\rm (b - y)}$ caused by a few stars with significant
reddening. In the light of these results, we assume that all of our
programme stars are unreddened and use the observed $b - y$ to determine
$T_{\rm eff}$, except for 5 stars beyond 100 pc with significant
reddening ($E{\rm (b - y)} \geq  0.025$) for which we use the
corrected $b - y$.\footnote{The stars are  HD\,15398, HD\,157467, HD\,159972,
HD\,163363, and HD\,213802.}

We use the calibration of Str\"{o}mgren
indices given by Alonso et al. (1996). This calibration, which uses
a large number of lower main sequence stars and subgiants whose temperatures were measured
by the infrared flux method, spans ranges of 4000 K $\leq$ $T_{\rm eff}$
$\leq 7000$ K and $-$2.5 $\leq$ [Fe/H] $\leq$ 0, and is well suited to the
programme stars. The calibration relates $T_{\rm eff}$ with $b - y$, $c_1$,
and [Fe/H], with $b - y$ making the major contribution to the
calibration and $c_1$, the gravity sensitive index, and [Fe/H] making 
minor contributions. To apply the calibration $b - y$ and $c_1$ were taken
from Hauck \& Mermilliod (1998), while [Fe/H] values were estimated
from  Str\"{o}mgren photometry (see below). (The effect of
reddening on $c_1$ is negligible.)
The error in the derived $T_{\rm eff}$ may come from different sources:
uncertainties in the Str\"{o}mgren photometry,  reddening, and the
calibration of the absolute flux in the infrared. Alonso et al. (1996)
estimated an  uncertainty of 1.5$\%$ (90 K) by taking into account
both the systematic and accidental errors in the calibration.

\subsubsection{The Surface Gravity}

The surface gravity of a programme star is estimated by comparing its
position in the colour-magnitude diagram with theoretical isochrones.
This comparison provided the stellar masses and radii and, thus, the
surface gravities. Isochrones were taken from Bertelli et al. (1994);
they span all required stellar masses and metallicities. Allende Prieto
\& Lambert (1999) have used these isochrones similarly to determine
fundamental parameters  for stars in the {\it Hipparcos} catalogue
within 100 pc.

Application of the method began by selecting the subset of isochrones
with a metallicity immediately below that of the star's photometric
metallicity. With this subset, we searched for those that reproduced
the observed $B-V$  and  $M_{\rm V}$ within the observational errors; 
$B-V$, $V$, and the parallax ($p$) were adopted
from the {\it Hipparcos} catalogue. Then, the different possible solutions,
corresponding to different masses and ages, were averaged to obtain mean values
for the stellar parameters and an estimate of the uncertainty from
the standard deviation.  This procedure was then repeated for a subset of
isochrones with a metallicity immediately above the star's
photometric metallicity, The observational error box for a given
star is defined by the uncertainty in the observed $B-V$ (which is taken
from the {\it Hipparcos} catalogue or put at 0.01 mag, whichever is larger),
and the $M_{\rm V}$,
as determined  from the 1-$\sigma$ errors in the parallax and $V$,

\begin{equation}
\sigma^2 (M_V) \simeq \sigma^2 (V) + 25 \frac{\sigma^2(p)}{p^2} \log^2 e 
\end{equation}

\noindent assuming $\sigma(V) = 0.07$ mag to force a minimum 
error in $M_{\rm V}$. Given that our sample satisfies
$\sigma(p)/p < 0.1$
we neglected the small bias in $M_{\rm V}$ and $\sigma(M_{\rm V)}$ introduced by
the non-linear dependence of $M_{\rm V}$ on the parallax (e.g. Brown et al. 1997).
In the majority of the cases, the error in $M_{\rm V}$
is dominated by the uncertainty in the 
parallax (as quoted in the {\it Hipparcos} catalogue).  
The estimated
final errors for log $g$ range from 0.03 to 0.10 dex.

\subsubsection{Photometric Metallicity}

The metallicity [M/H]\footnote {The photometric metallicity
[M/H] represents the elements
heavier than H and He, particularly iron peak elements
whose lines are numerous in the spectra} was determined from the $b-y$, $m_1$, and
$c_1$ indices using either Equation 2 (for F stars) or Equation 3
(for G stars) of Schuster \& Nissen (1989) with photometric
data from Hauck \& Mermilliod (1998). Using the
quoted uncertainties in (b-y), m$_{1}$, and c$_{1}$ from Hauck \& Mermilliod,
we estimate an uncertainty of $\simeq$ 0.2 dex in our photometric metallicity.
Hauck \& Mermilliod estimate a standard deviation of 0.16 dex in the 
[Fe/H] derived from their calibration.

\subsubsection{Microturbulence}

Earlier studies of the microturbulence in the atmospheres of
F and G dwarfs have shown that similar stars have very similar
levels of microturbulence which depends weakly on $T_{\rm eff}$
and $g$ (Nissen 1981). The microturbulence is determined
spectroscopically from the condition that the abundance derived from
lines of the same species should be independent of a line's
equivalent width. Often, Fe\,{\sc i} lines are used for obvious
reasons.

We determined $\xi_t$ for 87 of the 181 stars using 33 well defined
Fe\,{\sc i} lines with accurate $gf$-values (see below) and equivalent
widths of up to about 80 m\AA. Results are
well described by the relation

\begin{equation}
\xi_t = 1.28 + 3.3 \times 10^{-4}(T_{\rm eff} - 6000)
  -0.64(\log g - 4.5)
\end{equation}

where $\xi_t$ is in km s$^{-1}$, and $T_{\rm eff}$ and $g$ in their usual units.
This relation is derived for a sample of stars which have $T_{\rm eff}$ 
ranging from 5650~K to
6300~K, log $g$ ranging from 3.6 to 4.5, and  metallicity of $-$0.8 $\leq$ 
[Fe/H] $\leq$ 0.1.
The entire sample of our stars falls in the above range of $T_{\rm eff}$, 
log $g$, and
[Fe/H]. Thus, we can safely use the above derived relation for the rest 
of the stars 
in our sample expecting $\sigma$ $\approx$ 0.15\,km s$^{-1}$, which is 
the $rms$ error in the least-squares fit.

A similar linear regression has been used by others. Use of published
formulae results in slightly different results. For example, adoption
of EAGLNT's recipe gives a mean $\xi_t$ that is about 0.2 km s$^{-1}$
greater. Nissen's (1981) original expression  returns greater values but
by only 0.1 km s$^{-1}$. Chen et al. (2000) remark that their $\xi_t$
are 0.3 km s$^{-1}$ greater than EAGLNT's. 
Quite possibly, the lower values of $\xi_t$ found here are due to
our use of rather weak lines which are inherently less sensitive
to microturbulence. The difference is unimportant as far as the
abundance analysis is concerned; a change of $\xi_t$ by $\pm$0.25 km s$^{-1}$
changes the abundance of lines with equivalent widths of 50 m\AA\ or less
by less than 0.01 dex. Even at 100 m\AA\, the abundance changes by no more
than 0.04 dex.

\subsubsection{Comparison of Photometry and Spectroscopy}

As a check on the photometrically derived fundamental parameters
we used Fe\,{\sc i} and Fe\,{\sc ii} lines to determine $T_{\rm eff}$ and
log $g$ by the classical conditions that the Fe abundance be independent
of the lower excitation potential for Fe\,{\sc i} lines, and Fe\,{\sc i}
and Fe\,{\sc ii} lines yield the same abundance. Lines with reliable
$gf$-values (see below), and equivalent widths less than 60~m\AA\ were used,
a restriction that effectively eliminates
the sensitivity to the microturbulence. About 25 - 30 Fe\,{\sc i} and
4 Fe\,{\sc ii} lines were used.

Photometric and spectroscopic $T_{\rm eff}$'s  and metallicities [Fe/H] are compared in Figure\,2.
On average, spectroscopic temperatures are  hotter than their photometric
counterparts by 71 $\pm$ 47 K with a hint that the difference is $T_{\rm eff}$-dependent.
For [Fe/H], the mean difference between spectroscopic and photometric
estimates is merely 0.05 $\pm$ 0.09 dex with no detectable trend over the
range -0.2 to -0.8 in [Fe/H].

Surface gravity is checked using Fe\,{\sc i}-Fe\,{\sc ii}, 
and Cr\,{\sc i}-Cr\,{\sc ii}
lines. 
The comparison is made 
in Figure\,3. It is
seen that the neutral lines give a  slightly lower 
abundance: $\log\epsilon$(Fe\,{\sc i}) $-$ $\log\epsilon$(Fe\,{\sc ii}) = $-$0.02$\pm$0.05 with
just a hint of a dependence on [Fe/H]. Chromium gives a very similar result:
$\log\epsilon$(Cr\,{\sc i}) $-$ $\log\epsilon$(Cr\,{\sc ii}) = $-$0.04$\pm$0.06.
We conclude that the surface gravities do not introduce appreciable
systematic errors into the abundance analysis. 
Reducing the differences to exactly zero
calls for adjustments to the adopted atmospheric parameters that are within
their estimated uncertainties given above. Additionally, the negative
differences may signal departures from LTE effects, principally
the overionization (relative to LTE) of the neutral atoms (see, e.g., 
Trujillo Bueno \& Shchukina 2001).

\subsection{Stellar Ages}

We have estimated ages for the sample by comparison with the isochrones
published by Bertelli et al. (1994). As most of our stars have already evolved 
off the main sequence, we can constrain their age very precisely. For some
of the stars which are too close to the main sequence we can, at most, obtain
upper limits. Our method resembles that described by Lachaume et al. (1999).
Some aspects, however, are different, and therefore we describe it below.

We chose to use as observed quantities effective temperature and surface
gravity. The isochrones describe these parameters as functions of
the initial stellar mass $M_i$, the mass fraction of metals $Z$, and
the age $t$. A Gaussian distribution of relative errors 
was assumed for both parameters by adopting a probability density

\begin{eqnarray}
\begin{tabular}{cc}
$P (\log T_{\rm eff},\log g) \propto$ &
$\exp{ \left[ - \left( \frac{\displaystyle \log g - \log g^{*}}{\displaystyle \sqrt{2} ~\sigma(\log g)} \right)^2 \right] }$ \\
& \\
& $\exp{ \left[- \left( \frac{\displaystyle \log T_{\rm eff} - \log T_{\rm eff}^{*}}{\displaystyle \sqrt{2} ~\sigma(\log T_{\rm eff})} \right)^2 \right] }$, \\
\end{tabular}
\end{eqnarray}

\noindent which was used to determine the probability distribution for the age 

\begin{equation}
P(\log t) =  \int \int  ~  P (\log T_{\rm eff},\log g)  ~ dM_i ~ dZ.
\label{edad}
\end{equation}

\noindent  In practice, to find the best age
estimate for each star, we discretized the problem by sampling the isochrones
of Bertelli et al. with constant steps of 0.006 $M_{\odot}$ 
in the initial mass $M_i$, 0.05
in $\log t$ ($t$ in years), and 0.125 in $\log Z$. We then converted
the integral in Eq. \ref{edad} into a sum over the area 
confined by an ellipse centered at the 
adopted temperature and gravity (as listed in Table\,{1}), 
with semi-major axes three times the estimated  1-$\sigma$ uncertainties in 
$\log T_{\rm eff}$ and $\log g$. 
We imposed an additional constrain to the possible solutions by requiring
a metallicity within 0.25 dex from that spectroscopically determined.

Finally, the derived $P(\log t)$ was normalized and, whenever appropriate, fit 
with a Gaussian 
to derive the mean and a 1-$\sigma$ 
uncertainty for the age of the star, which are included in Table\,1. Figure\,4
shows an example of the practical application to two well-known nearby stars,
the Sun  and Procyon A. The upper panels show the position of the star
in the $\log T_{\rm eff}- \log g$ plane.  The dots are grid points within 
the 3-$\sigma$ error bar ellipse, which therefore were included
in the solution. In the lower panels, the probability density for the age
is displayed, as well as a best estimate and 1-$\sigma$ limits. 
Given the typical error bars for
our sample, a star with the solar parameters is too close to the ZAMS 
to determine but an upper limit. For a star with the parameters of 
Procyon A (see, e.g. Allende Prieto et al. 2002a), it is possible to constrain 
the age very precisely -- a case more representative of our sample 
regarding the evolutionary status.
The  ages derived in this study
show a range consistent with those published by Chen et al. (2000) and EAGLNT
for thin disc stars, but small differences are noticeable in Figure\,8.

\begin{table}
\centering
\caption{ Atmospheric parameters: $T_{\rm eff}$, log $g$, and [Fe/H],
and kinematic properties: $U_{\rm LSR}$, $V_{\rm LSR}$, $W_{\rm LSR}$, $R_{\rm m}$, $Z_{\rm max}$, $e$, and age
for the programme stars. See the appendix } 
\begin{tabular}{@{}llrl@{}}
\end{tabular}
\end{table}

\begin{table}
\centering
\caption{ Line data used in the analysis. Wavelength (\AA), ion, low-excitation potential (eV), $gf$-value, measured
solar equivalent width, and the references for the $gf$-values are given for each line.
{\bf \large { Table\,2 is available electronically.}}}
\begin{tabular}{@{}llrl@{}}
\end{tabular}
\end{table}

\begin{table*}
\centering
\begin{minipage}{110mm}
\caption{Selection of $gf$-values for Fe\,{\sc i} and Fe\,{\sc ii} lines.
For lines
lacking laboratory $gf$-values astrophysical $gf$-values are derived
by inverting solar and stellar spectra. Last column gives the adopted $gf$-values.}
\begin{tabular}{llccrcccl}
\hline \hline
Ion & W$_{\lambda}$  & LEP &\multicolumn{5}{c}{log $gf$ }&\\

    &($\AA$)    & (eV)    &  Oxford$^{a}$ & Hannover$^{b}$ & Solar & Star1  & Star2 &Adopted \\
\hline
Fe\,{\sc i}&5141.75    & 2.424 &        &     & $-$2.194 & $-$2.224 & $-$2.256& -2.205  \\
      &5247.06    & 0.087 & $-$4.946 &     &      &  & &-4.946 \\
      &5358.12    & 3.300 &        &     & $-$3.170 & $-$3.154 && -3.162    \\
      &5412.79    & 4.440 &        & $-$1.716 &      && &-1.716  \\
&5661.348   & 4.280 &        & $-$1.756 &&&&-1.756\\
&5778.458   & 2.590 & $-$3.475 & $-$3.430 &&&&-3.453\\
&5784.661   & 3.400 &        & $-$2.532 &&&&-2.532\\
&5809.220   & 3.884 & $-$1.609&      &&&&-1.609\\
&5849.690   & 3.695 & $-$2.935 &        &&&&-2.935\\
&5852.23    & 4.549 &        &        &  $-$1.181 &&&-1.161\\
&5855.090   & 4.608 &        &  $-$1.478 &&&&-1.478\\
&5856.10    & 4.294 &        &     &  $-$1.561& $-$1.545&&-1.548\\
&5858.79    & 4.220 &        &     &  $-$2.190& &&-2.180\\
&5859.60    & 4.550 &        &     &  $-$0.617& $-$0.581& $-$0.597&-0.588\\
&5862.37    & 4.550 &        &     &  $-$0.265& $-$0.302& $-$0.351&-0.293\\
&5956.700   & 0.859 & $-$4.605 &     &&&&-4.605\\
&6027.06    & 4.070 &        &     &  $-$1.167&&$-$1.104&-1.116\\
&6151.620   &2.176  &  $-$3.299& $-$3.265 & $-$-3.286&&&-3.282\\
&6159.38    & 4.610 &        &     &  $-$1.837& $-$1.841&$-$1.783&-1.820\\
&6165.36    & 4.143 &        &     &  $-$1.461& $-$1.452& $-$1.473&-1.455\\
&6173.340   & 2.223 & $-$2.880 &     &&&&-2.880\\
&6200.320   & 2.609 & $-$2.437 &     &&&&-2.437\\
&6213.44    & 2.223 &        &     &  $-$2.542& $-$2.600& $-$2.592&-2.558\\
&6240.652   & 2.220 &        &$-$3.233&&&&-3.233\\
&6265.141   & 2.176 &  $-$2.550&      &&&&-2.550\\
&6271.283   & 3.330 &        &$-$2.703&&&&-2.703\\
&6297.801   & 2.223 & $-$2.740 & $-$2.727 &&&&-2.734\\
&6322.694   & 2.588 & $-$2.426 &        &&&&-2.426\\
&6358.69    & 0.859 &        &     &         $-$4.056& $-$4.120&$-$4.222&-4.113\\
&6436.41    & 4.186 &        &     &         $-$2.364& $-$2.319&&-2.342 \\
&6481.878   & 2.279 & $-$2.972 &      &&&&-2.972\\
&6498.950   & 0.958& $-$4.699 &&&&&-4.699\\
&6518.374   & 2.830 &        & $-$2.450&&&&-2.450\\
&6574.233   & 0.990& $-$5.004 &&&&&-5.004\\
&6581.214   & 1.480 &        &  $-$4.680&&&&-4.680\\
&6591.33    & 4.593 &        &     &         $-$1.949&$-$1.873&&-1.911\\
&6608.04    & 2.279 &        &     &         $-$3.913&$-$3.929&$-$3.929&-3.924\\
&6625.027   & 1.010 & $-$5.336 &      &&&&-5.336\\
&6699.142   & 4.590 &        &  $-$2.101&&&&-2.101\\
&6713.75    & 4.795 &        &     &        $-$ 1.389& $-$1.367&$-$1.367&-1.374\\
&6725.36    & 4.103 &        &     &         $-$2.158& $-$2.149& $-$2.195&-2.167\\
&6739.524   & 1.560 &        &  $-$4.794&&&&-4.794\\
&6750.160   & 2.424 &   $-$2.621& $-$2.608&&&&-2.615 \\
&6752.711   & 4.640 &        &$-$1.204&&&&-1.204\\
&6837.009   & 4.590 &        &$-$1.687&&&&-1.687\\
&6733.15    & 4.638 &        &     &         $-$1.400&&&-1.390\\
&6857.25    & 4.076 &        &     &         $-$2.040&$-$2.025&&-2.028\\
&6971.936   & 3.020 &        &$-$3.340 &&&&-3.340\\
&7112.173   & 2.990 &        &$-$2.990 &&&&-2.990\\
&7751.12    & 4.990 &        &     &         $-$0.727&$-$0.704&$-$0.666&-0.692\\
&7802.51    & 5.080 &        &     &         $-$1.332&$-$1.294&$-$1.274&-1.300\\
&7807.92    & 4.990 &        &     &         $-$0.492&$-$0.499&$-$0.525&-0.499\\
&8365.644   & 3.250 &        & $-$2.037&&&&-2.037\\
&8757.200   & 2.845 &   $-$2.118&     &&&&-2.118\\
\end{tabular}
\end{minipage}
\end{table*}

\begin{table*}
\centering
\begin{minipage}{110mm}
\contcaption{}
\begin{tabular}{llccrcccl}
\hline \hline
Ion & W$_{\lambda}$  & LEP &\multicolumn{5}{c}{log $gf$ }&\\
    &($\AA$)    & (eV)    &  Oxford$^{a}$
  &  Hannover$^{b}$
  & Solar & Star1  & Star2 &Adopted \\
\hline
Fe\,{\sc ii} 5234.620 & 3.221 &    &                 & -2.22 &    &     & -2.22\\
&5425.26    & 3.200 &        &     &         $-$3.246&$-$3.187& $-$3.129&-3.177\\
& 6149.25    & 3.889 &        &     &         $-$2.713&$-$2.737& $-$2.611&-2.680\\
&6247.56    & 3.892 &        &     &         $-$2.341&$-$2.329& $-$2.204&-2.281\\
&6369.46    & 2.891 &        &     &         $-$4.100&$-$4.083& $-$4.043&-4.072\\
&6432.680   & 2.891 &        &$-$3.520 &&&&-3.520\\
&6456.39    & 3.903 &        &     &         $-$2.124&$-$2.146&&-2.115\\
&7479.700   & 3.892 &        &     &         $-$3.602&$-$3.640& $-$3.517&-3.586\\
&7515.840   & 3.903 &        &$-$3.407&      &&&-3.42$^{c}$ \\
\hline
\end{tabular}
\raggedright
$^{a}$ $gf$-values measured at Oxford (Blackwell et al. 1995 and 
references therein) \\
$^{b}$  $gf$-values measured at Hannover
(Bard et al. 1991; Bard \& Kock 1994)  \\
$^{c}$ Mean of $gf$-values measured
by Hannaford et al. (1992) and Hiese \& Kock (1990)
\end{minipage}
\end{table*}

\section{Atomic Data}

The critical atomic datum is the $gf$-value of a line. Our
general procedure was to search the literature for accurate theoretical
calculations or laboratory measurements of the $gf$-values. Our search left gaps
which were filled in various ways, including an inverted solar
analysis. 

References to the principal sources of data on $gf$-values for individual
elements are given
in Table\,2.
We comment here on the Fe\,{\sc i} and  Fe\,{\sc ii} $gf$-values because
these lines play a special role in the abundance analysis, as already
noted.
Our search uncovered accurate $gf$-values
for 33 of the 56 Fe\,{\sc i} and 2 of the 8 Fe\.{\sc ii} lines in our
lists of measured weak unblended stellar lines. In the case of the
Fe\,{\sc i} lines, these values
come mainly from three sources (see references in Table\,3). These and
other sources were reviewed by Lambert et al. (1996) who remarked on their
inter-agreement and suggested corrections to put all $gf$-values on a common
scale. 
To obtain $gf$-values for the remaining Fe\,{\sc i} and Fe\,{\sc ii} lines,
we calculated astrophysical $gf$-values from the {\it solar} (Iris) spectrum
using 
average abundances derived from Fe\,{\sc I} and Fe\,{\sc ii} lines
that have accurate theoretical or laboratory
$gf$ values.

We chose 9 Fe\,{\sc ii} lines that are unblended and
measurable in most stars in the sample. For 3 of the Fe\,{\sc ii} lines
(6369.46~\AA,
6432.68~\AA, and 7515.84~\AA)
laboratory measured $gf$-values are available. For the Fe\,{\sc ii} line at
6369.46~\AA\ Heise \& Kock (1990) measure log $gf$-value of $-$3.55 which
is significantly higher than our derived solar value of $-$4.10.
Our derived value is in fair agreement with the predicted log\,$gf$-value of
$-$4.25 (Kurucz 1998) and solar value of $-$4.36 (Blackwell, Shallis, \& Simmons 1980).
By adopting the laboratory $gf$-value, this line yields a
lower abundance than the mean Fe abundance (7.42) derived from
6432~\AA\ and 7515~\AA. The 6369.46~\AA\ Fe\,{\sc ii} line appears
to be blended with an other Fe\,{\sc ii} line at 6369.375\AA\ (Kurucz 1998),
but, the blend's contribution is negligible.
For this line, we adopted the $gf$-value derived in this study (Table\,3).
A search for laboratory  Fe\,{\sc ii} lines at $\lambda> 3000$ \AA\
has been presented by Allende Prieto et al. (2002a). They noticed that
Fe\,{\sc ii} 5234.6~\AA\ was measured by Kroll \& Kock (1987) 
and Heise \& Kock (1990). 
The re-scaled and averaged
value ($-2.23  \pm 0.08$) is in excellent agreement with our astrophysical 
determination ($-2.22$).

Astrophysical $gf$-values were also determined from the spectra of two of
the programme stars: HD\,145937 with [Fe/H] = $-$0.55, and HD\,217877 with
[Fe/H] = $-$0.14. The iron abundances and the model atmospheres were derived
using the Fe\,{\sc i} and 
Fe\,{\sc ii}
lines with accurate experimental $gf$-values. For the Fe\,{\sc i} and Fe\,{\sc ii} lines lacking
measured or accurate theoretical $gf$-values, we have computed them
by inverting the two stellar spectra, and using the mean Fe\,{\sc i} and
Fe\,{\sc ii} abundances, respectively. Finally, a mean of the solar and stellar
$gf$-values were adopted.
A similar procedure
is adopted for the rest of the elements. Results are in Table\,3.

In addition to the $gf$-values, hyperfine (hfs) and/or isotopic splitting must
be considered for a few lines. Lines of Sc\,{\sc ii},
V\,{\sc i}, and Co\,{\sc i}
are  broadened by hfs but test calculations
showed that the effect on the abundances is negligible for lines
of the observed equivalent widths. This is not the case for the Mn\,{\sc i}
and Cu\,{\sc i} lines. In these
cases, equivalent widths were calculated from synthetic line profiles  and
abundance found by matching these widths to the observed values.
For  the Mn\,{\sc i} lines, the data on the
splitting and strengths of the hfs components were taken from Kurucz (1998).
The lines show so much hfs that
without this detailed treatment they would return erroneous
abundances. In the case of the Cu\,{\sc i} lines, the 5105~\AA\ line
required thorough consideration of the hfs and isotopic splitting: the hfs 
data were
taken from Kurucz (1998) and the isotopic ratio was assumed to be the
solar system value for all stars ($^{63}$Cu/$^{65}$Cu = 2.24).
The effect of the hfs on the final abundances can be
as large as 0.6 dex for the stronger Mn lines where the
spacing between hfs components is large. In the case of
the 6021\,\AA\ Mn line and the 5218\,\AA\ Cu line, the separation between the
components is very small. 

\section{The Solar Abundances}

The spectrum of Iris for the Sun  was treated in the same way as the programme 
spectra, using an  ATLAS9 model for the accepted solar
parameters $T_{\rm eff}$ = 5780 K, $\log g = 4.44$, and a solar
composition ($\log \epsilon$(Fe) = 7.50 was adopted). Equivalent
widths were measured for the lines in Table\,2. The microturbulence
adopted was $\xi_t$ = 1.22 km s$^{-1}$, which
was derived using the Fe\,{\sc i} lines having accurate $gf$-values.
Analysis of the iron lines returned the parameters  $T_{\rm eff}$ = 5760$\pm50$ K,
$\log g = 4.44\pm0.10$ and $\log \epsilon$(Fe) = 7.45$\pm0.06$.
The parameters, $T_{\rm eff}$ and $\log g$ are within their uncertainty, equal to the
accepted values.
The derived solar Fe abundance $\log \epsilon$(Fe) =
7.45 $\pm$ 0.05 is close to the Grevesse \& Sauval photospheric value of
$\log \epsilon$(Fe) = 7.50 $\pm$ 0.05. The difference of 0.05 dex is
partly attributable to the 
use of different model atmospheres. Replacing ATLAS9
by Holweger-M\"{u}ller, the abundance increases to $\log \epsilon$(Fe) =
7.53. 

\begin{table}
\centering
\begin{minipage}{90mm}
\caption{The adopted solar abundances derived
by employing the Kurucz (1998) solar model in this study
are compared with the photospheric
solar abundances from the literature (Grevesse $\&$ Sauval 1998). The values in the brackets
are the standard deviations for the species represented by 3 or more lines.}
\begin{tabular}{lcllr}
\hline
\hline
Species &  no. of lines & \multicolumn{2}{c} {log $\epsilon$(X)} & Diff.  \\
        &               &  This study    & Literature &               \\
\hline
C~I       &       5       &  8.51 (0.06) & 8.39\footnote{Carbon and oxygen abundances
are taken from Allende Prieto et al. (2001b, 2002b)} &   0.12 \\
N~I       &       2       & 8.06  & 7.92 &  0.14 \\
O(7774) &       3       & 8.86 (0.05)   & 8.69$^{a}$ &   0.17 \\
\lbrack{O~I}\rbrack   &       1       & 8.73    & 8.69  &  0.04 \\
Na~I      &       2       & 6.27   & 6.33 &   -0.06 \\
Mg~I      &       3       & 7.54 ( 0.06)  & 7.58 &  -0.04 \\
Al~I      &       5       & 6.28 (0.05 )  & 6.47 &  -0.19 \\
Si~I    &       7       & 7.62 (0.05)   & 7.55 &  -0.07 \\
Si~II   &       1       & 7.64          & 7.55 &  -0.09 \\
S~I       &       3       & 7.34(0.09)     & 7.33  & 0.01 \\
K~I       &       1       & 5.22          & 5.12  & 0.10 \\
Ca~I      &       5       & 6.33 (0.07)   & 6.36  & -0.03\\
Sc~II   &       3       & 3.24 ( 0.14)  & 3.17 &  -0.07 \\
Ti~I      &       7       & 4.90 (0.06)   & 5.02  & -0.12 \\
V~I       &       6       & 3.93 (0.03)   & 4.00  & -0.07 \\
Cr~I    &       4       & 5.68 ( 0.07)  & 5.67 &   0.01 \\
Cr~II   &       1       & 5.65          & 5.67 &  -0.02 \\
Mn~I    &       3       & 5.37 (0.05)   & 5.39  & -0.02    \\
Fe~I    &      56       & 7.45 (0.04)   & 7.50 &  -0.05 \\
Fe~II   &       9       & 7.45 (0.07 )  & 7.50 &  -0.05 \\
Co~I      &       3       & 4.93 (0.04)   & 4.92  & 0.01 \\
Ni~I      &      18       & 6.23 (0.04)   & 6.25  & -0.02 \\
Cu~I      &       3       & 4.19 (0.05)   & 4.21  & -0.02 \\
Zn~I      &       2       & 4.47    & 4.60  & -0.13 \\
Sr~I      &       1       & 2.64          & 2.97  & -0.35 \\
Y~II    &       4      & 2.12 (0.04)   & 2.24 & -0.12 \\
Zr~II  &       1       &  2.45         & 2.60 &  -0.15 \\
Ba~II  &       3       & 2.20 (0.10)    & 2.13  & 0.07\\
Ce~II  &       3       & 1.58\footnote{The $gf$-values are derived by inverting the 
solar spectrum using the literature solar abundance}      & 1.58   & 0.00    \\
Nd~II  &       2       & 1.50$^{b}$           & 1.50  & 0.00 \\
Eu~II  &       1       & 0.61           & 0.51  & 0.10 \\
\hline
\end{tabular}
\end{minipage}
\end{table}

\begin{table}
\centering
\caption{ Elemental abundance for elements from C to Ti relative
to Fe for the programme stars. See the Table in appendix}
\begin{tabular}{@{}llrl@{}}
\end{tabular}
\end{table}

\begin{table}
\centering
\caption{ Elemental abundance for elements from V to Eu relative
to Fe for the programme stars. See the Table in Appendix}
\begin{tabular}{@{}llrl@{}}
\end{tabular}
\end{table}

Elemental abundances derived for the Sun from lines with
adopted $gf$-values, and equivalent widths measured from the {\it solar}
Iris spectrum (Table\,2)
are summarized in Table\,4. 
Agreement with the solar photospheric abundances
given by Grevesse \& Sauval (1998) was taken as
evidence of reliable $gf$-values.  
Their analysis uses an
empirical  solar model atmosphere and this difference in models introduces
small differences in the abundances derived from identical sets of atomic data.
In general, our results from a few
lines are in good agreement with standard results from usually more
lines.

\section{Stellar Abundances}

Elemental abundances for all the programme stars were determined using the
OVER ATLAS9 model computed for the $T_{\rm eff}$ and $\log g$ derived
from the photometry and {\it Hipparcos} parallaxes, respectively. Adopted model
parameters, and the $gf$-values are given in Table\,2. The stellar
abundance results
are referenced to the solar abundances determined in the current 
study (Table\,4) using the same lines, and
a similar procedure as for the programme stars, i.e.,
we derive and discuss differential abundances [X/H] and [X/Fe].
Abundances relative to iron for the entire sample are presented
in Table\,5 and Table\,6.
Before discussing astrophysical implications
of our results, we assess their accuracy, and present a few comparisons
with some recent analyses of F-G disc dwarfs.

\begin{table*}
\centering
\begin{minipage}{110mm}
\caption{Abundance uncertainties due to estimated uncertainties
in atmospheric parameters for five representative stars. 
The $\sigma$'s are quadratic sum of variations in
abundance ratios, [X/Fe], due to uncertainties in model parameters. The column $\sigma_{mod}$,
is the mean of $\sigma$'s and the quoted error $std$ is the standard deviation. }
\begin{tabular}{@{}lcccccc@{}}
\hline \hline
            & HD\,70  & HD\,94385  & HD\,9670 & HD\,6834 & HD\,112887  
 &   \\ 
$T_{\rm eff}$&  5649~K & 5814~K     & 6032~K   & 6290~K   & 6422~K                      \\
\lbrack Fe/H\rbrack & $-$0.5 & 0.01 & $-$0.30 & $-$0.70 & $-$0.30   &  \\
& $\sigma_{1}$ & $\sigma_{2}$& $\sigma_{3}$&$\sigma_{4}$& $\sigma_{5}$& $\sigma_{mod}\pm{std}$  \\
\hline 
\lbrack Fe/H\rbrack  & 0.09 & 0.08 & 0.08 & 0.06 & 0.06 & 0.07$\pm$0.01   \\ 
\lbrack C/Fe\rbrack  & 0.18 & 0.17 & 0.14 & 0.09 & 0.14 & 0.14$\pm$0.04   \\
\lbrack N/Fe\rbrack  & ...  & 0.20 & 0.17 & ...  & ...  & 0.19             \\
\lbrack O/Fe\rbrack  & 0.20 & 0.17 & 0.15 & 0.14 & 0.13 & 0.16$\pm$0.03   \\
\lbrack Na/Fe\rbrack & 0.04 & 0.03 & 0.04 & ...  & 0.02 & 0.03$\pm$0.01   \\
\lbrack Mg/Fe\rbrack & 0.01 & 0.06 & 0.02 & 0.04 & 0.02 & 0.03$\pm$0.02   \\
\lbrack Al/Fe\rbrack & 0.05 & 0.05 & 0.04 & 0.03 & 0.03 & 0.04$\pm$0.01   \\
\lbrack SiI/Fe\rbrack& 0.08 & 0.06 & 0.05 & 0.03 & 0.02 & 0.05$\pm$0.02   \\
\lbrack S/Fe\rbrack  & ...  & 0.08 & 0.08 & ...  & 0.11 & 0.09$\pm$0.02    \\
\lbrack K/Fe\rbrack  & 0.07 & 0.09 & 0.08 & ...  & ...  & 0.08$\pm$0.01   \\
\lbrack Ca/Fe\rbrack & 0.02 & 0.04 & 0.02 & 0.01 & 0.04 & 0.03$\pm$0.01   \\
\lbrack Sc/Fe\rbrack & 0.14 & 0.13 & 0.11 & 0.09 & 0.08 & 0.11$\pm$0.03   \\
\lbrack Ti/Fe\rbrack & 0.04 & 0.04 & 0.03 & 0.02 & 0.02 & 0.03$\pm$0.01  \\
\lbrack V/Fe\rbrack  & 0.05 & 0.04 & 0.03 & ...  & 0.03 & 0.04$\pm$0.01   \\
\lbrack Cr/Fe\rbrack & 0.02 & 0.03 & 0.01 & ...  & 0.00 & 0.02$\pm$0.01   \\
\lbrack Mn/Fe\rbrack & 0.03 & 0.06 & 0.02 & 0.01 & 0.01 & 0.03$\pm$0.02   \\
\lbrack Co/Fe\rbrack & 0.03 & 0.02 & 0.02 & ...  & 0.01 & 0.02$\pm$0.01   \\
\lbrack Ni/Fe\rbrack & 0.03 & 0.03 & 0.02 & 0.01 & 0.01 & 0.02$\pm$0.01   \\
\lbrack Cu/Fe\rbrack & 0.02 & 0.03 & 0.01 & 0.02 & 0.03 & 0.02$\pm$0.01   \\
\lbrack Zn/Fe\rbrack & 0.08 & 0.06 & 0.05 & 0.03 & 0.02 & 0.05$\pm$0.02   \\
\lbrack Sr/Fe\rbrack & 0.02 & 0.05 & 0.01 & ...  & ...  & 0.03$\pm$0.02   \\
\lbrack Y/Fe\rbrack  & 0.09 & 0.12 & 0.10 & 0.08 & 0.08 & 0.09$\pm$0.02   \\
\lbrack Zr/Fe\rbrack & ...  & 0.12 & 0.11 & ...  & 0.08 & 0.10$\pm$0.02   \\
\lbrack Ba/Fe\rbrack & 0.12 & 0.13 & 0.11 & 0.13 & 0.12 & 0.12$\pm$0.01   \\
\lbrack Ce/Fe\rbrack & 0.12 & 0.12 & 0.10 & ...  & 0.08 & 0.11$\pm$0.02   \\
\lbrack Nd/Fe\rbrack & ...  & 0.12 & 0.10 & ...  & 0.07 & 0.10$\pm$0.03   \\
\lbrack Eu/Fe\rbrack & 0.12 & 0.13 & 0.10 & ...  & 0.08 & 0.11$\pm$0.02   \\
\hline
\end{tabular}
\end{minipage}
\end{table*}

\subsection{Internal Assessment of the Errors}

This assessment of the errors afflicting the
derived abundances is made without questioning the
assumptions in the approach (i.e., LTE is accepted as valid).
In what is now a standard procedure, we calculate the
effect on the abundances of the errors in the
observed equivalent widths,  the defining model
atmosphere parameters, and the atomic data.

\subsubsection{Equivalent Widths}

On a typical spectrum, the accuracy of an equivalent width $W_\lambda$
is about 2~m\AA. This estimate arrived at from
independent attempts to measure a given line is roughly consistent
with the prescription given by Cayrel (1988):

\begin{equation}
\Delta W_\lambda = \frac{1.6\sqrt{w\delta x}}{S/N}
\end{equation}

where $w$ is the FWHM of the line, $\delta x$ is the pixel
size in \AA, and $S/N$ is the signal-to-noise ratio per pixel
in the continuum. For our spectra, $w$ is in the range
0.15~\AA\ to 0.25~\AA, $\delta x$ = 0.041~\AA, and $S/N \simeq 200$ to
400. These parameters give $\Delta W_\lambda$ $\simeq$ 0.4 m\AA\ to
0.9 m\AA. Given that the recipe does not consider blends and
the normalization to a local continuum, our adoption of a 2 m\AA\ limit
is reasonable.  For elements represented by more than one line,
there is a reduction of the abundance error arising from the
measurement error of the $W_\lambda$s. We suppose that the reduction 
scales as
$\sqrt{N}$, where $N$ is the number of lines, down to a minimum
value.

An external check is possible using our measurements of the Iris (solar)
spectrum with those from  lunar/day sky spectra by EAGLNT and Chen et al.
We have 47 lines in common with EAGLNT and 59 with Chen et al. Equivalent
widths are compared in Figure\,5 (upper panels).  The solar abundance
differences using a common model and atomic data are
shown in the lower panels of Figure\,5. EAGLNT's $W_\lambda$s are on average
1.4~m\AA\ smaller than ours.
Chen et al.'s measurements are close to ours: the mean difference in
$W_\lambda$ is just 0.3 m\AA.  
Our stellar spectra
are typically of the quality of the Iris spectrum. 

\subsubsection{Atmospheric Parameters}

In Section 3.3, we derived the atmospheric parameters and
discussed their uncertainties concluding that $\Delta T_{\rm eff}$ 
$\simeq$ $\pm$ 100 K, $\Delta\log g \simeq \pm 0.1$,
$\Delta\xi_t \simeq \pm 0.25$ km s$^{-1}$, and $\Delta$[M/H] $\simeq \pm 0.2$.
Assuming that the effects of these errors on the derived abundances
are uncorrelated for the small range of the various $\Delta$s, it
is a simple matter to determine their effect on the abundances.
We have done this for a sample of stars spanning the parameters range
of the entire sample. Predicted values of the total
abundance uncertainty are summarized in Table\,7. 

\subsection{External Errors}

Before making intercomparisons of our abundances with earlier studies, we
compare the abundances derived from the adopted Kurucz OVER models
with those derived from NOVER models. For 16 stars in our sample,
spanning  the full range of $T_{\rm eff}$ and [Fe/H] of the sample,
we recomputed abundance ratios [X/Fe] using the same atmospheric 
parameters, but
with NOVER models. 
The abundance differences from OVER and NOVER models are very small (see Figure\,6 for
representative elements). In the case of high excitation lines (see [C/Fe] and [O/Fe]),
there appears to be a trend of the abundance difference with [Fe/H] which is possibly
due to the fact that these excited lines are formed deep in the atmosphere where differences
between OVER and NOVER models are larger. Differences may be overestimated,
as we have used atmospheric parameters and $gf$-values derived from OVER models
in computing abundances from NOVER models. 

Chen et al.'s sample of 90 F-G dwarfs includes 23 from our collection.
Intercomparison of their and our results 
offers a  check on our results, but it must be deemed incomplete
in that the methods of analysis are very similar. 
First, we note that there is close agreement over the adopted
atmospheric parameters. In the sense Chen $-$ ours, we find
$\Delta T_{\rm eff} = 7 \pm$ 42 K, $\Delta\log g = 0.04 \pm 0.13$,
$\Delta$[Fe/H] = -0.02 $\pm 0.06$, and $\Delta\xi_t = 0.39 \pm 0.22$
km s$^{-1}$. This broad agreement reflects the similarity
in the approaches to the determination of atmospheric parameters.
For example, both studies use $(b - y)$ and Alonso et al.'s
calibration to obtain $T_{\rm eff}$.

Second, there is good agreement over the derived abundances for
elements in common. Both studies adopt a differential approach
using the solar spectrum.
There is partial overlap
in the lists of selected lines. Model grids differ: Chen et al. used
new MARCS models, and we used the OVER ATLAS9 models. 
In this initial comparison, we consider the mean differences
between the two studies. Later, we comment on a few
specific elements.
Mean abundance differences from the 23 stars are given in
Table\,8. These small zero-point differences likely arise from a
combination of factors: differences in the adopted solar equivalent widths,
use of different grids of model atmospheres for the Sun and the programme
stars, and selection of different lines with differing sensitivity to
the various atmospheric parameters. 
We apply the appropriate zero-point difference in cases where
we combine Chen et al.'s results with ours as a way to increase the sample size.

A direct comparison was made with EAGLNT's results for 3 stars
(HD\, 69897, HD\, 216385, and HD\, 218470) that are in common with
the present study.
EAGLNT's $T_{\rm eff}$ and log $g$
are found to be greater than our values by 96$\pm$41~K and 0.13$\pm$0.07 dex,
respectively. The difference in $T_{\rm eff}$ and log $g$ can possibly be attributed
to differences in adopted methods: EAGLNT determined
both the $T_{\rm eff}$ and log $g$ using 
Str\"{o}mgren indices. The differences in abundance
ratios, [X/Fe], are small ($<$ 0.1dex).
For a more reliable transformation of EAGLNT's abundances to our scale, we take
advantage of the fact that
Chen et al.'s selection of
stars included 26 from EAGLNT. The principal difference in the
adopted atmospheric parameters is in the effective temperatures for
which the difference in the sense Chen $-$ EAGLNT is -88 $\pm$ 56 K.
Other differences are quite minor: $\Delta\log g = -0.07 \pm 0.08$,
and $\Delta$[Fe/H] = $-0.02 \pm 0.07$. Abundance differences Chen $-$ EAGLNT
(see Table\,2 of Chen et al.) are small. 
The abundance differences Chen - ours and EAGLNT - ours are given in Table\,8.

\begin{table}
\centering
\caption{Average differences of atmospheric parameters and abundances for 24 stars
that are found to be common with Chen et al. (2000), and differences
between EAGLNT and ours (see the text for details). }
\begin{tabular}{@{}lrcrc@{}}
\hline \hline
Quantity     &  Chen-Ours  & $\sigma$ &   EAGLNT - Ours  & $\sigma$  \\
\hline
$T_{\rm {eff}}$(K)  &  7  & 42           &           77   &  41 \\
log $g$ (cm s$^{-2}$)         & 0.04 & 0.13         &           0.13 &  0.07 \\
$\xi_{\rm t}$(km s$^{-1}$)   & 0.39 & 0.22         &  0.19   &   --- \\
\lbrack Fe\,{\sc i}/H \rbrack & $-$0.02 & 0.06 &  $-$0.01 & 0.04 \\ 
\lbrack Fe\,{\sc ii}/H \rbrack & 0.01 & 0.09   &   0.01 & 0.03 \\ 
\lbrack Na\,{\sc i}/Fe \rbrack & $-$0.09 & 0.06 &  $-$0.01 & 0.03  \\ 
\lbrack Mg\,{\sc i}/Fe \rbrack & 0.06 & 0.04    &  0.08 & 0.09 \\ 
\lbrack Al\,{\sc i}/Fe \rbrack & $-$0.11 & 0.07 & $-$0.07 & 0.08  \\ 
\lbrack Si\,{\sc i}/Fe \rbrack & 0.01 & 0.03    &  0.03 & 0.02 \\ 
\lbrack K\,{\sc i}/Fe \rbrack & $-$0.05 & 0.05  &  ...  & ... \\ 
\lbrack Ca\,{\sc i}/Fe \rbrack & 0.06 & 0.03    &  0.01 & 0.02 \\ 
\lbrack Ti\,{\sc i}/Fe \rbrack & 0.01 & 0.07    &  0.11 & 0.09  \\ 
\lbrack V\,{\sc i}/Fe \rbrack & 0.03 & 0.07     &  ...  &  ... \\ 
\lbrack Cr\,{\sc i}/Fe \rbrack & 0.03 & 0.05    &  ...  &  ...\\ 
\lbrack Ni\,{\sc i}/Fe \rbrack & 0.04 & 0.02    &  0.04 & 0.03 \\ 
\lbrack Ba\,{\sc ii}/Fe \rbrack & 0.06 & 0.08   &  0.03 & 0.02 \\ 
\hline
\end{tabular}
\end{table}

\section{Chemical Evolution of the Disc}

Several signatures of chemical evolution of the Galactic
disc may be looked for using our data.  Here, we comment
briefly on the age-metallicity relation, aspects of the
evolution of the relative abundances (i.e., [X/Fe] versus [Fe/H]),
the scatter in relative abundances at a fixed [Fe/H], and the
difference in compositions of thick and thin
disc stars.

Our sample is nearly homogeneous being comprised of
thin disc stars to the almost total exclusion of thick
disc representatives (see below for our definition
of the thick disc). Furthermore, our sample was
selected to cover only a part of the [Fe/H] range
spanned by thin disc stars.  As appropriate, we
combine our data with published results.

\subsection{Age-Metallicity Relation}

EAGLNT's  age-metallicity relation which hinted at a slow
drop in metallicity with increasing age of the star
was marked by a large spread in metallicity at a fixed age.
Our sample taken at face value offers a cleaner relation --
see Figure\,7 where the relation is shown for [Ca/H], [Fe/H], and
[Ba/H]. Figure\,8 shows this relation with earlier results
from Chen et al. and EAGLNT with their marked scatter in [Fe/H]
at a fixed age.
The appearance of a cleaner age-metallicity relation is
attributable to two selection effects. First and more
important, we chose
stars in a restricted [Fe/H] range and, in particular,
[Fe/H] $>$ $-$0.2 are poorly represented. Second, our sample
is kinematically homogeneous to the  almost complete
exclusion of thick disc stars.
In short, we support
earlier conclusions that the age-metallicity relation
offered directly by local thin and thick disc stars is characterized by
a large scatter about a slow monotonic decrease of metallicity
with increasing age.

\subsection{Relative Abundances}

In discussing the variation of [X/Fe] versus [Fe/H] (Figures~9, 10 \& 11), we begin
by making brief comparisons with published results,
principally those from the recent 
extensive surveys of disc F and G dwarfs by EAGLNT, Feltzing \&
Gustafsson (1998), Fulbright (2000), and Chen et al. over the common interval
in [Fe/H]. 
Note that our stars sample well the interval
[Fe/H] $\simeq$  $-$0.1 to $-$0.6. EAGLNT's stars covered a
slightly broader range ([Fe/H] = $-$0.8 to $+$0.2) with
Feltzing \& Gustafsson adding metal-rich stars ([Fe/H]
= 0.0 to $+$0.2).
EAGLNT considered O, Na, Mg, Al, Si, Ti, Fe, Ni, Y, Zr, Ba, and Nd.
Chen et al. provided good coverage for
[Fe/H] = $-$1.0  to $+$0.1 for O, Na, Mg, Al, Si, K, Ca, Ti, V,
Cr, Ni, and Ba.
Fulbright (2000) analysed a sample of disc and halo stars with a wide range in 
$T_{\rm eff}$, log $g$, and metallicity, and provided abundances
for 13 elements (Na, Mg, Al, Si, Ca, Ti, V, Cr, Ni, Y, Zr, Ba, and Eu).
From his study we have selected 77 disc dwarfs/subgiants that have $-$1.0 $\geq$ [M/H] $\leq$ +0.2, 
and 3.8 $\leq$ log $g$ $\leq$ 5.0.
For elements (C, N, S, Sc, Mn, Co, Cu, Zn, and Ce) not covered by these
surveys, we compare with other
studies. 

For many elements, inspection of the plots of [X/Fe] versus [Fe/H] shows
that  published results differ from ours by only  
a small zero-point correction. Table\,8 lists the zero-point
corrections between Chen et al. and us, and 
EAGLNT and us.
Since we have only 3 stars in common with EAGLNT, we
infer the mean zero-point correction between our and EAGLNT's results
by combining the corrections between Chen et al. and us, and Chen et al. and EAGLNT.
Each of the three analyses is a differential analysis
conducted relative to the Solar spectrum.
Given that 
theories of Galactic chemical
evolution should not pretend to challenge observations at the 0.1 dex
level, we shall not attempt to pin down the origins of the
zero-point corrections; possible sources of a zero-point difference
were discussed above. For applications where the largest possible 
sample size may be useful, we shall apply the
zero-point correction to place published results on our system of abundances.

Our results are in good agreement for elements common
with EAGLNT.
The trends of [X/Fe] versus [Fe/H] 
are identical for O, Na, Mg, Al, Si, Ca, Ti, Ni, and Ba. Differences may
be noticeable for Y, and Zr. In the case of Y and Zr, we find [X/Fe] to
decline slightly with decreasing [Fe/H], but EAGLNT found either
no (Y) or an opposite trend (Zr). The most striking differences with
respect to EAGLNT are not in the trends but in the scatter
about a mean trend. Most notably, the scatter found here is considerably
smaller than reported by EAGLNT for Mg, Al,  and Ti, especially at [Fe/H] $\leq$ $-$0.4. The 
scatter in
[X/Fe] for these and other elements is discussed below.

There is also very good agreement with Chen et al.'s results.
After allowing for a small  zero-point difference, 
two differences are noted. 
First, there is a small but distinct difference between our (and EAGLNT's)
and Chen et al.'s run of [Al/Fe] versus [Fe/H]: we find
[Al/Fe] to increase slightly with decreasing [Fe/H], but Chen
et al. find an initial decrease from [Al/Fe] $\simeq$ $+$ 0.15
at [Fe/H] = 0 to [Al/Fe] $\simeq$ $-$0.1 at [Fe/H] $\simeq$ $-$0.4
from which point [Al/Fe] may increase slightly.
Second, there are elements for which the analyses agree about the
trends but give different results for the scatter in [X/Fe] at a fixed
[Fe/H]: notably, Cr for which Chen et al. report a flat trend but with
stars spanning the range [Cr/Fe] $\simeq$ $-$0.1 to $+$0.2 in contrast to
our results (Figure\,10).

Remarks on [X/Fe] versus [Fe/H] are offered for those elements not
covered by either EAGLNT or Chen et al. Our reference is usually to the
most recent work  on an element. The elements in question are
C, N, S, Sc, Mn, Co, Cu, Zn, Ce, and Eu. Oxygen is also
discussed.

Carbon: Our results are in good agreement with abundances derived by
Gustafsson et al. (1999) from the 8727 \AA\ [C\,{\sc i}] line in 80
stars of EAGLNT's sample. We confirm the increase of [C/Fe] with
decreasing [Fe/H]. Scatter about the mean trend is larger from
our C\,{\sc i} lines than from the forbidden line, a difference
attributed to the different sensitivities of the lines to the
atmospheric parameters. 
The slight offset - our [C/Fe] are larger by
about 0.03 dex - is in line with the systematic difference  in the adopted
$T_{\rm eff}$s. 

Tomkin et al.'s (1995) analysis of C\,{\sc i} lines in 105 of EAGLNT's stars
gave
a very similar slope for the [C/Fe] -- [Fe/H] relation as found by us and
by Gustafsson et al. but Tomkin et al.'s results are offset to lower [C/Fe]
by about 0.15 dex from ours. This offset arises because EAGLNT's $T_{\rm eff}$
scale is about 80 K hotter than ours.  

Nitrogen: Atomic nitrogen is represented in spectra of F-G dwarfs by
weak N\,{\sc i} lines.
Our [N/Fe] values show considerable scatter at a fixed [Fe/H], which
is largely attributable to the weakness of the two N\,{\sc i}
lines, and the $T_{\rm eff}$ sensitivity of the derived abundances.
At those metallicities ([Fe/H] $\simeq -$0.2) well represented in our
sample, [N/Fe] is about 0.2, which implies that, if [N/Fe] = 0 at [Fe/H]
=0, [N/Fe] increases as [Fe/H] falls below the solar value.
The N\,{\sc i} lines are not detectable in our stars with
[Fe/H] $\leq -$0.4.

A selection of N\,{\sc i} lines was used previously by
Clegg, Lambert, \& Tomkin (1981) to measure [N/Fe] in 15 stars over the
[Fe/H] spanned by our stars to give [N/Fe] $\simeq$ 0.0, a value smaller
than ours by about 0.2 dex.
Considering the similar sensitivities of the C\,{\sc i} and N\,{\sc i}
lines to the adopted atmospheric parameters, [N/C] is robustly
model-independent. Our result [N/C] $\simeq$ 0 is consistent with
that given by Clegg et al.

Oxygen: 
The forbidden [O\,{\sc i}] 6300~\AA\ line
in our spectra falls near the gap between the spectral orders and the line can
be measured reliably only in stars where lines are blue-shifted.
This and the weakness of the line limited us to measure
the 6300~\AA\ line in 60 stars ranging in metallicity from 0.0 to $-$0.5~dex.
In analyzing the $W_{\lambda}$ of 6300~\AA\ we have considered the blend caused
by a Ni\,{\sc i} line at 6300.335~\AA\ with the $gf$-value
 reported by Allende Prieto, Lambert \& Asplund (2001b).
The effect of the blend on the [O\,{\sc i}] abundance is significant, especially
in stars of solar-metallicity. The abundance ratio [O/Fe] is found
to increase with decreasing [Fe/H], in general agreement with other studies
where the blend was recognized (Nissen et al. 2002).

Oxygen abundances also come from the relatively strong 
infrared triplet at 7775~\AA. The strength
of these lines allowed us to derive oxygen abundance for all the stars in our sample. 
Direct LTE
analysis of these lines give [O/Fe] values noticeably greater than from the [O\,{\sc i}]
6300~\AA\ line, a discrepancy attributed to non-LTE effects on the
permitted lines (Kiselman 1993). 
In order to extend the [O/Fe] results to lower
[Fe/H] than possible with the [O\,{\sc i}] line, we derive
the following empirical correlation from stars for which O\,{\sc i} and [O\,{\sc i}]
lines are analysed:
\vskip 0.3cm
\noindent
${\rm {[O/Fe]_{6300}}} = {\rm {[O/Fe]_{7775}}} + 0.1138 \times {\rm { [Fe/H]}} - 0.5425 \times {\rm log} T_{\rm eff}$
\begin{equation} 
  +0.0925 {\rm log} g + 1.4891
\end{equation}

with an $rms$ scatter of 0.11 dex.
This equation is used to correct the [O/Fe] from the 7775~\AA\ to the scale
of the
forbidden line. Our final results for [O/Fe] are shown in Figure\,9. 

Direct comparisons with the majority of the 
published analyses of the [O\,{\sc i}] lines are compromised
by their neglect of the Ni\,{\sc i} blending line. 
Our results are in fair
agreement with 
Nissen et al. (2002).

Sulphur: 
Recently, Chen et al. (2002) reported
S abundances for a sample of 26 disc stars.
Our current results, based on 3 S\,{\sc i}
lines that are common with Chen et al (2002), in general, agree
with their results. However, the scatter  appears higher in our data.
A difference between Chen et al. (2002) and our analysis
is the [S/Fe] offset from [S/Fe] = 0 even at [Fe/H] $\sim$ 0. We discuss
below the possible reasons for the small offsets.

Potassium: In spite of a large scatter, a weak trend of increasing K abundance
with decreasing [Fe/H] is noticeable (see Fig\,9). 

Scandium: Nissen et al. (2000) obtained Sc abundances for stars 
in the Chen et al. sample. These were revised by Prochaska \& McWilliam
(2000) using accurate hyperfine splittings for the Sc\,{\sc ii} lines.
The revised results in the common [Fe/H] interval are in
good agreement with ours. Inspection of Figure\,10 shows that Sc, as noted
by Nissen et al. but disputed by Prochaska \& McWilliam, behaves
like the $\alpha$-elements Ca and Ti.
This similarity may not
hold for stars with [Fe/H] $< -$1 (Prochaska \& McWilliam 2000; Gratton \&
Sneden 1991).

Manganese: A decrease of the Mn abundance (relative to Fe) for decreasing
[Fe/H] was
noted by Beynon (1978), explored with high quality spectra of
a few stars by Gratton (1989), and defined for F-G disc dwarfs by
Nissen et al. (2000) and Prochaska \& McWilliam (2000). For [Fe/H] 
$> -$0.6, the limit of our sample, the [Mn/Fe] measured by us and by
Prochaska \& McWilliam are in good agreement as to slope and
absolute value. 

Cobalt: Data on cobalt in F-G disc dwarfs is especially sparse.
Gratton \& Sneden (1991) obtained [Co/Fe] $\simeq$ $-$0.1 from
a few stars in our [Fe/H] range. This early result is
consistent with ours.

Copper: Our results show little variation of [Cu/Fe] with [Fe/H]. Earlier
discussions have focussed on the run of [Cu/Fe] with [Fe/H] for
halo stars for which [Cu/Fe] declines steadily for decreasing [Fe/H]
(Sneden, Gratton,
\& Crocker 1991): [Cu/Fe] = 0.38[Fe/H] $+$0.15. This slope is at odds
with our results, but close inspection of the few results available to
Sneden et al. for stars with [Fe/H] $>$ $-$1 shows that a change of
slope might have been anticipated.

Zinc: This element appears to behave similarly to the $\alpha$-elements;
[Zn/Fe] increases slightly with decreasing [Fe/H]. This result is
consistent with Sneden et al.'s (1991) analyses of Zn in disc and
halo dwarfs and giants which gave [Zn/Fe]
constant (= 0.04) from [Fe/H] = $-$0.1 to $-$2.9.

Cerium: In solar material, Ce is principally an $s$-process product.
Thus, it is the expected result that [Ce/Fe] and [Ba/Fe] vary in
very similar fashions. That the scatter at a given [Fe/H] is
larger for [Ba/Fe] than [Ce/H] is probably due (see below) to the
fact that the Ba\,{\sc ii} but not the Ce\,{\sc ii} lines are
strong, and so dependent on the adopted microturbulence and other
factors. 

Neodymium: Extending the argument just given for Ce, we note that
the $s$-process contributions to solar abundances of Ba, Ce, Nd, and
Eu are 81, 77, 56, and 6 per cent, respectively. Noting the
opposite slopes of [Ba/Fe] and [Eu/Fe] versus [Fe/H], it is not
surprising that [Nd/Fe] appears to be independent of [Fe/H].

Europium: Recently, Eu abundances for samples of EAGLNT's stars
using their atmospheres were published by Woolf, Tomkin, \& Lambert
(1995) and Koch \& Edvardsson (2002). Their results for [Eu/Fe]
are consistent with ours.

\subsection{Offsets in [X/Fe]}

Given the differential analysis and the absence of cosmic
scatter in our sample (see next section), one would expect stars of solar [Fe/H]
to have [X/Fe] =0. 
However, Figures\,9, 10, \& 11 show [X/Fe] of Mg, Al, Si, S offset
by 0.03 to 0.05 dex at [Fe/H] = 0.0, and [X/Fe] of elements Sc, V, Mn, Cu
offset by $\sim$0.05 at [Fe/H] =0.0. 
A key to the origins of the small offsets may be the fact that the Sun
is not fully representative of our sample. With $T_{\rm eff}$ =5780~K, log $g$ = 4.44, and [Fe/H] = 0,
the Sun is in a tail of the stellar distribution in all three parameters (Figure\,1). This
has several consequences. The equivalent widths of some lines in the stellar spectra may
differ considerably from their strengths in the solar spectrum. Systematic errors such as those arising
from the neglect of stellar granulation and departures from LTE which would cancel
in a strictly differential analysis of quite similar stars may leave a small
imprint here in the form of the offset in [X/Fe] at [Fe/H]=0. A leading
NLTE effect may be overionization of the metals in the iron-group. One might
suppose similar degrees of overionization such that [Ni/Fe] may 
be reliably
estimated from a combination of Ni\,{\sc i} and Fe\,{\sc i} lines. 
In the case of Mg, Si, and S, the
offset disappears when [X/Fe] is computed using iron abundance 
derived from Fe\,{\sc ii} lines.

\begin{table*}
\centering
\begin{minipage}{140mm}
\caption{ Heavy element abundances for $s$-process enhanced outliers.}
\begin{tabular}{@{}lrrrrrr@{}}
\hline \hline
[X/Fe] & HD\,21922  & HD\,36667 & HD\,80218 & HD\,88446 & HD\,140324 & HD\,220842 \\
\hline
\lbrack Fe/H\rbrack  & $-$0.48  & $-$0.45 & $-$0.28 & $-$0.52 & $-$0.36 & $-$0.31 \\ 
\lbrack Sr/Fe\rbrack & 0.25    & 0.33  & 0.33 & 0.57 & ... & 0.27 \\
\lbrack Y/Fe\rbrack  & 0.23    & 0.38  & 0.26 & 0.58 & 0.29 & 0.35 \\
\lbrack Zr/Fe\rbrack & 0.18    & 0.33  & 0.16 & 0.65 & ... &0.27 \\
\lbrack Ba/Fe\rbrack & 0.33    & 0.47  & 0.00 & 0.66 & 0.18 & 0.09 \\
\lbrack Ce/Fe\rbrack & 0.34    & 0.26  & $-$0.07 & 0.58 & 0.29 & 0.05 \\
\lbrack Nd/Fe\rbrack & 0.17    & 0.40  & 0.04  & 0.69 & ... & 0.12 \\
\hline
\end{tabular}
\end{minipage}
\end{table*}

\subsection{Outliers}
There are few stars in our sample whose abundances for one or more
elements differ significantly from the
mean abundances of the rest of the stars at the same [Fe/H] (see Figures~9, 10 \& 11).
Two stars, HD\,110989 and HD\,136925, are enhanced only
in abundances of Mg, Al, Si, S, and Ti, and the rest of the abundances
are normal. The kinematics of these stars suggest they belong to the 
thick disc population, for which such enhancements are characteristic (see below).
Figures~9, 10 and 11 also reveal that half a dozen stars are enhanced
in $s$-process elements, i.e., Sr, Y, Zr, Ba, Ce, and Nd. Abundances
of these stars are summarized in Table\,9. HD\,88446
is known to be a $s$-process enriched CH sub-giant (Smith, Coleman \& Lambert 1993).
The current abundances are in very good agreement with the literature values.
We assume that
the other stars are also $s$-enriched dwarfs. The higher $s$-process
abundances in unevolved stars are attributable to mass-transfer from a
companion asymptotic giant branch (AGB) star that is now an unseen white dwarf.

\subsection{Cosmic Scatter?}
A first step towards an understanding of the scatter in the abundance ratios [X/Fe]
is to quantify the scatter. To do this
we fit a linear relation to the
data and examine the residuals about the relation. A histogram
of the residuals is fitted with a Gaussian. In very few cases is a
Gaussian  not a  good fit.  Figure\,12
shows [X/Fe] versus [Fe/H] for X = Mg, and Si with the fitted
linear relation,  the residuals about this relation, and a histogram
of these residuals with its fitted Gaussian.
The $\sigma_{\rm gau}$ of the Gaussian varies
from element to element. Table\,10  summarizes the results.
Note that in computing the measurement errors $\sigma_{\rm mod}$ for individual stars, 
we have not included the uncertainty in the EW measurements.
Except for two elements K and Sr, abundances are computed using 2 or more lines and the net random error
due to the EW measurement is very small. 

\begin{table}
\centering
\caption{ The predicted uncertainty, $\sigma_{mod}$, is compared
with $\sigma_{gau}$ of the Gaussian fit to the residuals}
\begin{tabular}{@{}lrr@{}}
\hline \hline
 \lbrack X/Fe\rbrack     &  $\sigma_{\rm mod}$  & $\sigma_{\rm gau}$ \\
\hline
\lbrack Fe/H\rbrack   & 0.07 & ... \\ 
\lbrack C/Fe\rbrack   & 0.14 & 0.07 \\
\lbrack N/Fe\rbrack   & 0.19 & 0.10  \\
\lbrack O/Fe\rbrack   & 0.16 & 0.07 \\
\lbrack Na/Fe\rbrack  & 0.03 & 0.04 \\
\lbrack Mg/Fe\rbrack  & 0.04 & 0.04 \\
\lbrack Al/Fe\rbrack  & 0.04 & 0.05 \\ 
\lbrack Si/Fe\rbrack  & 0.05 & 0.04 \\
\lbrack S/Fe\rbrack   & 0.09 & 0.09 \\
\lbrack K/Fe\rbrack   & 0.08 & 0.10 \\
\lbrack Ca/Fe\rbrack  & 0.03 & 0.04 \\
\lbrack Sc/Fe\rbrack  & 0.11 & 0.05 \\
\lbrack Ti/Fe\rbrack  & 0.03 & 0.04 \\
\lbrack V/Fe\rbrack   & 0.04 & 0.04 \\
\lbrack Cr/Fe\rbrack  & 0.02 & 0.03 \\
\lbrack Mn/Fe\rbrack  & 0.04 & 0.04 \\ 
\lbrack Co/Fe\rbrack  & 0.02 & 0.04 \\
\lbrack Ni/Fe\rbrack  & 0.02 & 0.03 \\
\lbrack Cu/Fe\rbrack  & 0.02 & 0.06 \\
\lbrack Zn/Fe\rbrack  & 0.05 & 0.06 \\
\lbrack Sr/Fe\rbrack  & 0.04 & 0.08 \\
\lbrack Y/Fe\rbrack   & 0.09 & 0.07 \\
\lbrack Zr/Fe\lbrack   & 0.10 & 0.07 \\
\lbrack Ba/Fe\rbrack  & 0.12 & 0.08 \\
\lbrack Ce/Fe\rbrack  & 0.11 & 0.08 \\
\lbrack Nd/Fe\rbrack  & 0.10 & 0.07 \\
\lbrack Eu/Fe\rbrack  & 0.11 & 0.08 \\
\hline
\end{tabular}
\end{table}

An obvious question occurs - is there information about Galactic chemical evolution (GCE) in the
$\sigma_{\rm gau}$ estimates, or do abundance measurement errors dominate? 
To address this, we compare the error $\sigma_{mod}$ previously calculated (Table\,10) with $\sigma_{gau}$.

Values of $\sigma_{gau}$ are well matched to the estimates of $\sigma_{mod}$
for almost every element (Figure\,13). This is especially true for Cr and Ni, two elements
spectroscopically similar to Fe with similar
nucleosynthetic origins (i.e., intrinsic star-to-star differences in [X/Fe] are minimized).
For C and N, $\sigma_{gau}$ is less than the estimated $\sigma_{mod}$ which suggests the
latter are overestimates. In these cases where the abundance is based on high excitation
lines, we suspect the $T_{\rm eff}$ errors are significantly overestimated: the
adopted error is likely an upper limit (see Alonso et al. 1996) to the
combination of the random and systematic errors, but only the latter is a
weak contributor to $\sigma_{mod}$. In the case of Cu, and Sr, $\sigma_{gau}$
exceeds $\sigma_{mod}$. Abundance of Sr is dependent on a single line and it is, therefore
difficult to argue that there is a real star-to-star scatter in [Sr/Fe].

The implication for GCE is clear: the intrinsic or cosmic scatter
in [X/Fe] among these thin disc stars with birthplaces concentrated at
galactocentric distances of 7 to 10 kpc is small, say
$\sigma_{cosmic}$ $<$ $\sigma_{gau}$. Noting
that the sampled ejecta come in differing proportions
from the three principal sites of stellar
nucleosynthesis -- SN~II, SN~Ia, and AGB stars -- the lack of
cosmic scatter implies that the ejecta from the different sites were
well mixed into the gas from which the stars formed. 

That there is cosmic scatter was suggested by EAGLNT's study which
showed a noticeably larger scatter in [Mg/Fe], [Al/Fe], and [Ti/Fe]
(relative to similar elements such as Si and Ca) for stars with [Fe/H] $\leq$ $-$0.4. \footnote{
EAGLNT's suggestion that some metal-rich stars were enhanced in Na, Mg, and Al was not
confirmed by Tomkin et al. 1997).} Reconciliation of this suggestion
with the evident lack of comparable scatter in our sample is attempted below.

\subsection{Chemical Evolution of the Thin Disc}

Chemical evolution as portrayed by Figures\,9, 10, and 11 is
broadly interpreted as the consequence of mixing into the
interstellar medium of ejecta from three principal sites of
nucleosynthesis: Type II SN, Type Ia SN, and AGB stars.
Qualitatively, the key features of
Figures\,9, 10, and 11 are widely accepted as understood. For example,
the gradual decline in [X/Fe] for $\alpha$-elements (O, Mg, Si, Ca, and Ti)
with increasing [Fe/H] is taken to reflect the  delayed
contribution from Type Ia SN with a lower $\alpha$/Fe ratio
than the  ejecta from Type II SN which dominated the
chemical evolution at earlier times and, hence, low [Fe/H].
Detailed modeling has been attempted by many authors --
see, for example,:
Timmes, Woosley, \& Weaver (1995), Chiappin et al. (1997), Goswami \&
Prantzos (2000), and Alib\'{e}s et al. (2001).
Quantitative matching of predictions to observations of [X/Fe]
remains elusive. 

Contributions from the AGB stars must be reflected in the
abundances of C, N, and heavy elements synthesized
predominantly by the $s$-process. Carbon and nitrogen
are also synthesized by massive stars. These stars may also
contribute Sr, Y, and Zr through the weak $s$-process and Eu
through the $r$-process.
While AGB stars also synthesize
these `light' elements, they dominate synthesis of
`heavy' $s$-process nuclides such that, as noted above,
AGB stars control production of Ba, Ce, and Nd 
in  thin disc stars .
In contrast, europium is an $r$-process or Type II SN
product.

Inspection of Figure\,11 shows, as mentioned in Section 7.2,
that Sr, Y, Zr, Ba, and Ce show very similar declines
in [X/Fe] with decreasing [Fe/H]. Europium shows
an increase, as expected of this $r$-process product.
Increasing Ba and Ce (relative to Fe) surely indicates
the increasing prominence of AGB star ejecta relative
to Type II and Type Ia ejecta which contribute Fe but very
little Ba  and Ce.
The ratio of heavy to light $s$-process abundances is
shown in Figure\,14.
The similarity of the slopes of [X/Fe]
versus [Fe/H] for  light (Sr, Y,  Zr) and heavy (Ba, Ce)
elements may reflect either  unchanging relative contributions
from AGB stars or a change in these contributions
which is offset by a change in the weak $s$-process
contribution from the
massive stars.  
On the assumption that the AGB
stars are the controlling influence, the unchanging abundance
ratio of
heavy to light elements indicates that the exposure to
neutrons in the $s$-process site is essentially independent
of the metallicity of the contributing AGB stars.

\subsection{Thin and Thick Discs}

The thick disc is  considered by many to be the result
of heating of the thin disc by accretion of, or merger with
small stellar systems. 
Freeman \& Bland-Hawthorn (2002) refer to the resulting thick disc
as a `` snap frozen '' relic of the state of the (heated) early disc.
The labels `` thin '' and `` thick '' were introduced
by Gilmore $\&$ Reid (1983) and denote the different vertical scale heights of the
populations: 300 pc for the thin disc and 1450 pc for the thick disc. The ratio
of stellar density of thick to thin disc is a few per cent near the Sun. 
The thick disc is broadly described as slightly metal-poor and old
relative to the thin disc.

In our sample, the lack of cosmic scatter in [X/Fe] at a fixed [Fe/H] is striking
and in apparent conflict with results obtained by others. We noted
above the contrast with EAGLNT's results for Mg, Al, and Ti where a larger
scatter- apparently, cosmic - was reported for stars with [Fe/H] less than
about $-$0.4 and more metal-rich than about $-$0.8; the lower bound
is uncertain due to the paucity of very metal-poor stars in their sample. Fuhrmann (1998),
in a notable contribution, attributed the scatter in [Mg/Fe] to a mixing of thin with
thick disc stars and a different chemical evolution of these two stellar
populations (see also Gratton et al. 2000 for remarks on O, and Mg with respect to Fe).
He suggested that at [Fe/H] $\approx$ $-$0.4 the [Mg/Fe] was either `high' or `low' with
no stars having an intermediate value. 

Prochaska et al. (2000) report results from the initial phases of a survey
of thick disc main sequence stars in the solar neighborhood. Stars
are selected to have V$_{LSR}$ from $-20$ to $-100$ km s$^{-1}$, 
metallicities in the
interval $-0.4$ to $-1.1$, and a W$_{LSR}$ that takes a star to at least 600 pc 
above the Galactic plane. The latter criterion in particular
takes out thin disc stars. A color selection provides stars generally cooler than those
comprising our and other surveys.  Ten stars were analysed using model 
atmospheres and high resolution spectra, obtaining abundances
for 20 elements. Presently, the sample lacks stars in common with other
surveys, and, hence, there may be small offsets between the [X/Fe] and those
of the surveys considered here. Results\footnote {We adopt
the values based on solar astrophysical $gf$-values, denoted [X/Fe]$^{S}_{n}$ by Prochaska et al.} are
incorporated into Figures\,15,16 and 18. 

Given the large stellar sample which may be assembled from various
sources, we explore in greater detail the evidence for cosmic scatter
in [Mg/Fe] and other elements arising from the mixing at a fixed [Fe/H]
of thin and thick disc stars.
Figure\,15 shows [Mg/Fe] and [Ti/Fe] versus [Fe/H] for
stars drawn from the samples referenced in the caption.
With the exception of the data from Prochaska et al.
(2000) and Fulbright (2000), the published abundances have been adjusted
to our scale  using small corrections. Common stars
among the samples were identified, and treated them only once:
for common stars in our study and others, we have adopted our values,
and for common stars in Chen et al. and EAGLNT, we have adopted
Chen et al. values.
The appearance of cosmic scatter at [Fe/H] $\simeq$
$-$0.4 is intimately related to the mixing of stellar
populations. For [Fe/H] $> -0.4$, the stars belong to the
thin disc. At [Fe/H] $< -0.4$, thick disc stars comprise
the majority.
In simplest terms stars with the higher [Mg/Fe] at a given
[Fe/H] are old thick disc stars from the inner Galaxy. To justify this
identification, we divide the stars into two groups according to whether [Mg/Fe]
is greater than or less than $+$0.2 dex as shown in Figure\,15. 

In the four panels of Figure\,16, we identify a star
by its group membership in plots of V$_{LSR}$, W$_{LSR}$, R$_{m}$, and age. We have
calculated these quantities when authors of the selected samples did not provide them.
Low and high [Mg/Fe] stars occupy almost non-overlapping areas of the panels
showing V$_{LSR}$, R$_{m}$, and age. Many  stars of high [Mg/Fe]
(large symbols in the plots) have a circular velocity less than that of the Sun, a range
in W$_{LSR}$ that shows they make excursions to about 1 kpc above the plane, an age
that places them among the oldest stars in the Galaxy 
(log$_{\tau_{9}}$$\approx$ 1.0 to 1.3),
and an origin in the inner Galaxy (R$_{m} \sim$ 5 to 7 kpc). 
The high [Mg/Fe] stars
are primarily representatives of the Galaxy's thick disc. The low [Mg/Fe] stars of the
same [Fe/H] as the thick disc stars are predominantly thin disc stars from galactocentric
distances closer to the radius of the Sun's orbit. 
This difference in populations
is, as Fuhrmann recognized, responsible for appearance of cosmic scatter in [Mg/Fe] among
slightly metal-poor stars. The difference
between two disc populations is also clear from Figure\,17, where
the stars are binned in 0.5~kpc intervals of $R_{\rm m}$. 
For this
we have considered star samples of EAGLNT, Chen et al., Fulbright (2000), 
and ours, totalling
around 500 disc stars. The entire sample is grouped into two classes: 
stars with [Mg/Fe] $\geq$ 0.2
as thick disc stars, and stars with [Mg/Fe] $<$ 0.2 dex as thin disc stars. Thin disc stars fit a Gaussian centering at 
$R_{\rm m}$ = 8.1 kpc and thick disc stars peak at $R_{\rm m}$ = 6.7 kpc.
The distribution 
suggests that thin disc stars are likely to have circular orbits and thick
disc stars are relatavely eccentric and formed 
closer to the Galactic centre.

The dispersion in W$_{LSR}$ for thin disc stars
increases progressively for lower [Fe/H] is in agreement with the correlation 
between $\sigma_W$ and age determined by G\'omez et al. (1997) who studied
several thousand B- to F-type stars with {\it Hipparcos} data, and the
[Fe/H] vs. age relationship depicted in Figure 8. G\'omez et al. found the
$\sigma_W$-age relationship to saturate at $\sim 4-5$ Gyr, which is also
consistent with the scatter in $W$ flattening for thin disc stars with [Fe/H] 
$\leq -0.3$ in Figure\,16.

\begin{table}
\centering
\caption{ Mean abundance ratios,[X/Fe] and kinematic properties
for thick and thin disc stars. Thin disc stars are taken
from present study with [Fe/H] from $-$0.35 to $-$0.70. Thick disc
values are for stars whose metallicity is $-$0.70 $\geq$ [Fe/H] $\leq$ $-$0.35 and
the kinematic values as defined in Figure\,18 and in this table. Thick disc stars
are from EAGLNT, Prochaska et al., and Woolf et al. }
\begin{tabular}{@{}lrrrrr@{}}
\hline \hline
Quantity     &  Thin disc  &Thick disc & $\Delta$[X/Fe]\\
\hline
$V_{\rm LSR}$ (Km s$^{-1}$) & $+$50 to $-$40 & $-$40 to $-$100 & \\  
$W_{\rm LSR}$ (Km s$^{-1}$) & $+$40 to $-$40 & $+$80 to $-$80& \\
$R_{\rm m}$  (kpc)         & $>$7.0         & 5.5 to 7.0 &  \\
log $\tau_{9}$ (yrs)       & $<$ 1.0        & $\geq$ 1.0 &  \\
\hline
\lbrack Na/Fe\rbrack  & 0.07$\pm$0.04 & 0.10$\pm$0.04 & 0.03 \\
\lbrack Mg/Fe\rbrack  & 0.09$\pm$0.05 & 0.30$\pm$0.07 & 0.21\\
\lbrack Al/Fe\rbrack  & 0.11$\pm$0.06 & 0.27$\pm$0.08 & 0.16\\ 
\lbrack Si/Fe\rbrack  & 0.07$\pm$0.04 & 0.18$\pm$0.05 & 0.11\\
\lbrack Ca/Fe\rbrack  & 0.05$\pm$0.04 & 0.16$\pm$0.06 & 0.11 \\
\lbrack Sc/Fe\rbrack  & 0.04$\pm$0.07 & 0.15$\pm$0.05 & 0.11\\
\lbrack Ti/Fe\rbrack  & 0.05$\pm$0.05 & 0.29$\pm$0.06 & 0.24\\
\lbrack V/Fe\rbrack   & 0.01$\pm$0.13 & 0.22$\pm$0.06 & 0.21\\
\lbrack Cr/Fe\rbrack  & -0.02$\pm$0.03 & 0.02$\pm$0.01& -0.04 \\
\lbrack Mn/Fe\rbrack  & -0.16$\pm$0.05 & -0.13$\pm$0.05 & 0.03 \\ 
\lbrack Ni/Fe\rbrack  & -0.02$\pm$0.02 & 0.03$\pm$0.02 & 0.05\\
\lbrack Ba/Fe\rbrack  & -0.02$\pm$0.15 & -0.10$\pm$0.06 & -0.08 \\
\lbrack Eu/Fe\rbrack  & 0.12$\pm$0.10 & 0.30$\pm$0.08  & 0.18\\
\hline
\end{tabular}
\end{table}

To quantify the differences between the thin and thick disc samples, we choose a sample of
thick disc stars as having [Fe/H] $\leq$ $-$0.35 and a V$_{LSR}$ from $-$40 to $-$100 km s$^{-1}$,
R$_{m}$ of 5.5 to 7.0 kpc, and an age log $\tau_{9}$ $\geq$ 1.0. The upper bound
chosen for [Fe/H] is seemingly the maximum [Fe/H] of thick disc stars seen in the
solar neighbourhood; we are not aware of selection effects compromising
the considered surveys which would have lead to the exclusion of stars of high [Mg/Fe] with
[Fe/H] $>$ $-$0.3. The lower bound on [Fe/H] for thick disc stars is unknown and unimportant
to our measurement of the difference in [X/Fe] between the two samples. Our comparison sample
of thin disc stars has V$_{LSR}$ between -40 and $+$40 km s$^{-1}$. Mean values for the
thin disc stars are $R_{m}$ = 8.1 kpc, and log $\tau_{9}$ = 0.6 for those stars with [Fe/H] between
$-$0.35 and $-$0.7, the metallicity range for which present samples of thin and thick disc
stars have an overlap.

Thin and thick disc stars with [Fe/H] between -0.35 and -0.7
differ in [X/Fe] (Figure\,18). Table\,11 gives the mean [X/Fe]
for those elements well represented in both samples and
the difference $\Delta[X/Fe] = [X/Fe]_{thick} - [X/Fe]_{thin}$
between the mean values.
Our $\Delta$-values are smilar to those suggested by
Prochaska et al. from a comparison between their [X/Fe]
and those of published surveys of disc stars.
$\Delta[X/Fe]$ is positive for
several elements: Mg, Al, Si, Ca, Sc, Ti, V, and Eu. To
within the uncertainty, $\Delta[X/Fe]$ is zero for Na, Cr,
Mn, and Ni, and possibly negative for Ba. Particularly striking
is the contrast between the odd-$Z$ light elements Na and Al,
a difference noted by EAGLNT. The $\Delta$-values for Sc and
V are largely dependent on Prochaska et al.'s results for
thick disc stars and may be affected by a systematic
offset arising from different abundance analysis techniques.
We note in Figure~18 that our and Chen et al's results
suggest a smaller $\Delta$ value for V.

Thick disc stars have a narrow spread in [Ba/Eu]. The spread
in [Eu/Fe] evident from Figure\,18 is much reduced when [Ba/Eu] is considered.
Figure\,19 shows this. A few thin disc stars share the [Ba/Eu] of the
thick disc stars. For a pure $r$-process solar like mix, the [Ba/Eu] $\approx$ -0.7.
Evidently, the thick disc stars have a smaller fraction of $s$-process heavy elements
than disc stars of the same [Fe/H]. This is an expected result given that
the thick disc stars are older than the thin disc companion stars; the $s$-process
elements are contributed by the more slowly evolving low mass AGB stars.

Our primary aim here was to establish that previous reports of cosmic
scatter in [Mg/Fe] and similar ratios arise from the mixing of stellar
populations. Cosmic scatter, as shown here, is undetectable in [X/Fe] at fixed [Fe/H]
among local thin disc stars. It is apparently small for thick disc stars but a
large sample subjected to a uniform analysis must be made available to test
this suspicion. 

Our selection of thick disk stars by negative $V_{LSR}$, high  $\tau_9$, low 
[Fe/H], and small $R_m$
excludes  two interesting groups of stars. First, there are stars with
[Mg/Fe] $>$ 0.2 with positive  $V_{LSR}$,  and low [Fe/H] (Figure\,16). Second,
there are a few stars with $V_{LSR}$ characteristic of the thick
disk but higher [Fe/H].

The first group by virtue of their generally positive $V_{LSR}$
appears  to originate from outside the solar circle ($R_m \simeq
8.5$ to 10 kpc).  They share the lower [Fe/H] of the thick
stars; none are present with [Fe/H] $\geq$ $-$0.3.
Although the sample size is small, there is a hint that, in
contrast to the thick stars, these stars are not exclusively old.
The  amplitude in $W_{LSR}$ across the 
sample appears to be smaller than that of the thick disk stars.
However, it is interesting to note that abundance pattern
of this group is very similar to that of thick disk stars as in Figure\,18, and Table\,11.
This group appears to be related to the thick disk stars discussed above with
$R_m$ of 5 to 7 kpc.

The second group are apparently schizophrenic. Additional stars
belonging to this group are to be found in Felzing \& Gustafsson (1998). Their  $V_{LSR}$ ($\sim - 50$ km s$^{-1}$)
would identify them as belonging to the thick disk. But by their $W_{LSR}$, $R_{\rm m}$,
spread in ages, and their [Fe/H] they would be linked with the thin
disc stars. These stars do not show the [X/Fe] of the thick disc stars.

\section{Concluding Remarks}

Our stellar sample is comprised primarily of thin disc stars, but by
drawing on samples which include thick disc stars, we may compare and
contrast chemical compositions of thin and thick disc stars.

\subsection{Thin Disc -- Questions}

Three aspects of the compositions of the thin disc stars attract
our attention: (i) the lack of cosmic scatter in [X/Fe], (ii) the
dispersion in the age-metallicity relation, and (iii) the origin of
the gas from which the most metal-poor thin disc stars were
formed.

The thin disc, as sampled by stars which are passing through the
solar neighbourhood, has a lower limit of [Fe/H] $\sim -$0.7, an
upper bound to the age of about 10 Gyr, and origins from Galactocentric
distances of between about 7 and 10 kpc. 
The restriction to $7<R_{\rm m}<10$ likely emerges as a direct consequence of the
small eccentricity of the orbits.
Between these limits
on  the metallicity  and
age, variations in [X/Fe] are small (Figures~9, 10, \& 11).
A more remarkable result concerning the thin disc is the confirmation 
of earlier work,
showing that, although the
thin disc stars sample a range in metallicity, age, and originate from
different Galactocentric distances,  these identifying characteristics
vanish almost entirely when relative abundances of elements
are considered. In particular, apart from weak trends with
[Fe/H], [X/Fe] exhibits no
scatter in excess of that attributable to measurement
errors.\footnote{A few stars are evidently $s$-process enriched
and  presumably result from mass-transfer from a now deceased
companion.}
This fact stands as a challenge to models of Galactic chemical evolution.

Lack of
scatter in [X/Fe] suggests that the Galactic thin disc is 
chemically homogeneous at a given time, and   mixing of the various ejecta
into star-forming clouds in the disc is very efficient.
This homogeneity refers to abundance ratios [X/Fe] not to the
abundances [Fe/H] or [X/H]. As noted in the discussion of the age-metallicity
relation, there is a spread in the latter quantities at a given age and
a given galactocentric distance.

If the mild evolution of [X/Fe] with [Fe/H] is ignored, [X/Fe]
is independent of the birthplace ($R_m$) and birthdate of a
thin disc star,
always bearing in mind that we are sampling a small range in $R_m$.
This independence extends to the present time.
Massive stars being young and, therefore, observed at or very close to their
birthplace may be used to trace the present composition of the Galaxy's thin
disc.
Observations
show that [X/Fe] $\simeq$ 0.0 irrespective of position in the Galaxy.
This result is well shown by Andrievsky et al.'s (2002a,b) analyses
of Cepheids with locations corresponding to Galactocentric distances of
4 to 10 kpc. (These authors assume the Sun to be at 7.9 kpc.)
The mean [Fe/H] of the Cepheids at the Sun's Galactocentric distance
is not significantly different from zero. The authors suggest that the
spread of about $\pm$0.15 dex
in [Fe/H] at this and other well sampled distances exceeds the
measurement errors.  A spread is reported from observations of
young open clusters (Friel 1995; Edvardsson 2002).
Iron abundance decreases slightly with increasing Galactocentric distance:
Andrievsky et al. obtain a slope for [Fe/H] of $-0.029 \pm0.004$ dex
kpc$^{-1}$, a value of the same sign but slightly smaller than a majority
of earlier measurements of this gradient. Other metals show rather
similar gradients except that the heavy elements La, Ce, Nd, Eu, and Gd
show a positive gradient (mean value of $+0.013$ dex kpc$^{-1}$).
Extension of the abundance analyses to Cepheids  in the
inner Galaxy shows that higher metallicity prevails inside about 6 kpc
(Andrievsky et al. 2002b): [Fe/H] $\simeq$ +0.3 from five stars at
Galactocentric distances of 4.4 to 5.7 kpc, but [X/Fe] $\simeq$ 0.0
with the exception of a few elements represented by very few lines.
In short, the Cepheids,  young stars  sampling a wide range in
Galactocentric distances, have a composition expressed as [X/Fe]
not distinctly different
from that of   thin disc stars.
Examination of Andrievsky et al.'s results shows that
the spread in [X/Fe] is small and likely dominated by
measurement errors.

Existence of a (weak) abundance gradient is pertinent to the
question of scatter in the age-metallicity relation.
Most determinations put the slope at a larger value than
the above value from Cepheids, say $-$0.1 dex kpc$^{-1}$ is typical.
Grenon (1987, 1989) has adduced evidence that the gradient was of a similar
magnitude in the past.  EAGLNT's analysis offered supporting
evidence. Presence of a shallow gradient is now widely linked to the
presence of a stellar bar in the inner Galactic disc. Observationally,
it is found that spiral galaxies with bars have flatter gradients
than those without bars (Martin \& Roy 1994). Theoretically,
bars have been shown to homogenize the gas (Martinet \& Friedli 1997).
Recently, Cole \& Weinberg (2002) have argued that the Galactic
bar is younger than 6 Gyr, which, if the bar were the only mechanism
for homogenizing the gas, would seem to imply the possibility of
a different abundance gradient at very early times. However, there
are other postulated ideas for maintaining a flat abundance
gradient (see Andrievsky et al. 2002a).

Attention was drawn to the large spread in the age-metallicity
relation by EAGLNT. At a fixed age, there is a
spread of about 0.5 dex in [Fe/H], or at a fixed [Fe/H] there is a
range in ages over about 8 Gyr.  If stars
migrate in Galactocentric distance and a Galactic abundance gradient
existed over part of the time sampled by the age-metallicity relation,
scatter in that relation would result.
Feltzing, Holmberg, \& Hurley (2001) from their analysis of nearly
6000 stars argue that stellar migration cannot account in
full for the scatter, and interpret the scatter as `intrinsic to the
formation processes of stars'.
One might wonder if a contributing
factor is an inability to predict accurately the Galactocentric
distance R$_m$, especially for stars from the inner Galaxy which
reach the Sun on `hot' orbits thanks to the action of the
inner bar of the Galaxy (Sparke \& Sellwood 1987; Raboud et al. 1998).

Discovery of the earliest manifestation of the thin disc would provide
an important datum about the disc's history. According to our
sample, the oldest stars (Figure~16) are 
slightly younger than  the average thick disc star. These thin
disc  stars extend in
[Fe/H] to $-$0.7 with, on average, a birthplace outside the
solar circle. The vertical velocities $W_{LSR}$  are
those of the thin disc and not the thick disc.
The lower bound to [Fe/H] was imposed as a selection criterion by
EAGLNT, Chen et  al., and ourselves. There are stars known at
lower [Fe/H] with thin disc kinematics (cf. Chiba \& Beers 2000; Beers et al. 2002).
Clearly, an important task is to measure compositions for a selection
of these metal-poor stars, and, in particular, to determine accurately
those [X/Fe] which distinguish thin from thick disc stars. As important
will be  tracing the evolution of [X/Fe] with [Fe/H] and finding where
(or if) [X/Fe] merges with halo values.\footnote{There are
halo stars ($V_{\rm LSR} \simeq -200$ km s$^{-1}$  with some [X/Fe] similar to those
of the thin disc at [Fe/H] $\sim -0.6$. These are
the so-called $\alpha$-poor stars found by Nissen \& Schuster (1997).
Disc and $\alpha$-poor stars have similar [X/Fe] for O, Mg, Si, Ca, Ti, Cr, Y, and Ba, but
not for Na and Ni, which are underabundant (relative to Fe) in
the $\alpha$-poor stars. Thick disc (and other halo) stars analysed
by Nissen \& Schuster have [X/Fe] like those of other
thick disc stars, a similarity that excludes the possibility of
systematic errors causing the Na and Ni abundance anomalies.
It is difficult to see how to link the $\alpha$-poor halo stars to
a widespread property of thin disc stars. Nissen \& Schuster
supposed that these halo stars have been accreted from a dwarf
galaxy with a history of nucleosynthesis different from ours.}

\subsection{Thick Disc  -- Questions}
Thick disc stars selected by $V_{LSR}$ relative
to thin disc stars of the same [Fe/H] are older, exhibit a large dispersion in
$W_{LSR}$, and originate from smaller Galactocentric distances
(Figure~16). 
Given the larger eccentricity and velocity dispersion of the thick disc 
stars compared to the thin disc population, we would expect a larger
range in R$_m$. If $\rho_{\rm thick}$(R) is nearly as flat as 
$\rho_{\rm thin}$(R), Figure\,17 would reveal a wider Gaussian for the thick disc, 
but still centred at R$_{\odot}$. The fact that the distribution of thick disc
stars is shifted to smaller mean galactocentric distances can be
explained by assuming a steeper dependence of $\rho_{\rm thick}$
with galactocentric distance. 
Furthermore, the shift could be interpreted as the direct observation of
the truncation radius of the thick disc, which , as pointed out by Freeman
\& Bland-Hawthorn, may be different from the thin disc's, marking the size
of the thin disc when the thick disc formed.

As we noted already, we confirm and extend an earlier
result (Fuhrmann 1998; Gratton et al. 2000;
Prochaska et al. 2000)
that there are  differences in [X/Fe] between thick
and thin disc stars of the same [Fe/H]. These
differences are summarized in Table\,11. Especially, notable
is the different behaviour of Na, Mg, and Al (relative to Fe) between the
thick and thin disc. In particular, as first stressed
by EAGLNT, [Na/Mg] at a fixed [Mg/H] is lower by about 0.2 dex
for the thick relative to thin disc stars but [Al/Mg]  is the
same for both groups.
These differences are clues to the origins of the populations, and
to the nucleosynthesis of Na,Mg, Al, and other elements.
The reader is referred to Prochaska et al. (2000) for discussion of these
points.

Following Fuhrmann (1998), one may harbour the view that
distribution functions for [X/Fe] are non-overlapping for samples
of thin and thick disc stars for elements such as Mg for which
$\Delta[X/Fe]$ is large. Quantitative spectroscopy of
a large sample of thick disc stars and a control sample of
thin disc stars is now needed to establish the distribution functions.
Thick disc stars at the low [Fe/H] limit discussed here appear to merge
with bulge/halo stars as far as [X/Fe] is concerned.
Feltzing \& Gustafsson's (1998) analysis of metal-rich F-G stars  shows no
change in [X/Fe] with $V_{LSR}$, and so hints that the thick - thin
disc differences disappear by [Fe/H] $\simeq$ 0.0, but the low $W_{LSR}$ of their stars
imply that their stars do not travel far from the Galactic plane, and, hence, attribution
to the thick disc may be questionable.

Prochaska et al. discuss at some length the implications for the
evolution of the Galaxy of the different
[X/Fe] in thin and thick disc stars.
As long noted, the extension of
the halo [X/Fe] for $\alpha$-elements to higher [Fe/H] in
thick disc than thin disc stars is suggestive of a delay in the contribution
made by SN Ia to the thick disc. Certainly, the thick-thin differences
in [X/Fe] raise interesting questions about the nucleosynthesis by
SN II. A key point stressed by Prochaska et al. concerns the
closely similar ages and [X/Fe] for bulge, thick-disc, and (the majority of the)
halo stars which suggests that these populations formed from the same
gas at about the same time. Several speculations may be offered
to account for the lack of thin disc stars with thick disc
abundances: (i)
the gas from which thick disc stars formed did not contribute to the thin disc;
(ii) a delay in star formation in the thin disc
enables SN Ia to enrich the gas in Fe-group elements and so reduce
the Mg/Fe for the first generation of thin disc stars; (iii)
perhaps, the thick disc gas was diluted with gas in the thin disc
before the observed stars formed.

\subsection{New Challenges}
New challenges to the observer are presented by our survey of thin
disc stars and the comparisons with published analyses of thick disc
stars. A particular challenge is to find and 
survey the composition of thick disc
stars over more of the space defined
by $R_{\rm m}$, $\tau_{\rm a}$, $W_{\rm LSR}$, and [Fe/H].
Given the availability of high-resolution spectrographs
and large telescopes and tools for standard  abundance
analysis, it should not be difficult to provide the compositions. 

A different challenge should be noted: the need to step beyond
a  classical abundance analysis with its reliance on the
classical model atmosphere and method of line analysis. One
step is taken by replacing the assumption of LTE by non-LTE
(i.e., statistical equilibrium) in analysing the absorption
lines. A more challenging step involves using 3-dimensional
hydrodynamic model atmospheres in place of the classical
atmosphere which adopts hydrostatic equilibrium among its
defining assumptions. Ultimately, the combination of the
hydrodynamical models with non-LTE is desired.
Consummation of this marriage may be necessary in order
to detect and quantify the cosmic scatter in the abundance
ratios [X/Fe], and to obtain finally definitive results
for [X/Fe].

\section{Acknowledgments}
We thank Bengt Edvardsson for providing a list of stars with Str\"{o}mgren photometry,
from which the current sample is taken, and  the code for calculating the
U,V,W space motion components. Johannes Andersen and Stephane Udry for providing the CORAVEL
radial velocities prior to publication. We thank Jon Fulbright for making available
computer code, with the kind permission of D. Lin, to calculate orbital parameters.
We thank David Yong, Gajendra Pandey, and Nils Ryde for many useful discussions.
This research has been supported in part by National Science Foundation 
(grants AST 96-18414, AST 99-00846, AST 00-86321)
and the Robert A. Welch Foundation of Houston, Texas.
This research has made use of the SIMBAD
data base, operated at CDS, Strasbourg, France, and the NASA ADS, USA.

{}

\clearpage
\begin{figure*}
\epsfxsize=18truecm
\epsffile{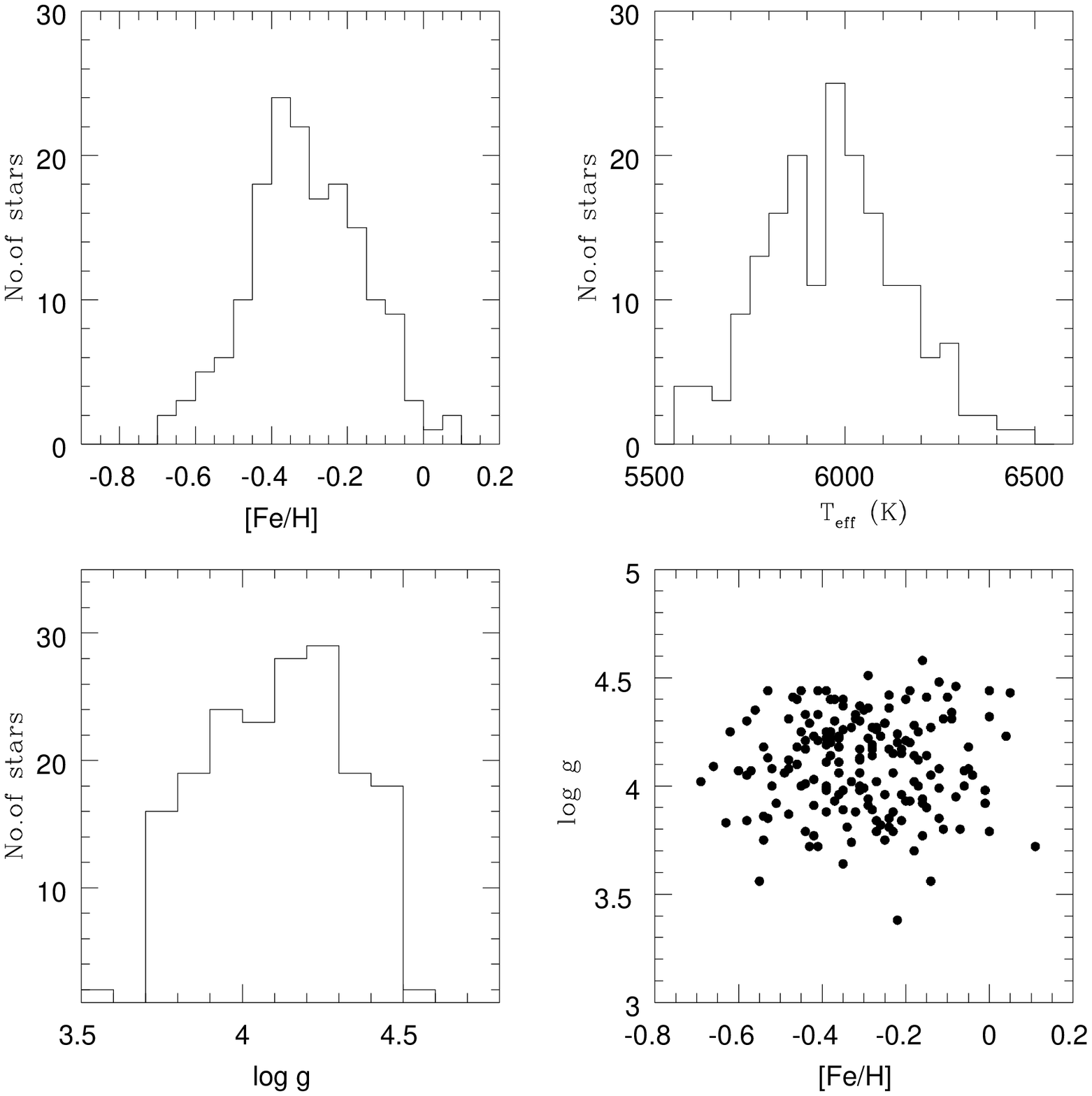}
\caption{Distribution of our stellar sample in [Fe/H], $T_{\rm eff}$, and gravity. Also shown is 
the gravity versus metallicity.}
\end{figure*}

\clearpage
\begin{figure*}
\centering
\includegraphics[bb=5.0cm 5cm 15cm 22.0cm,width=9.6cm]{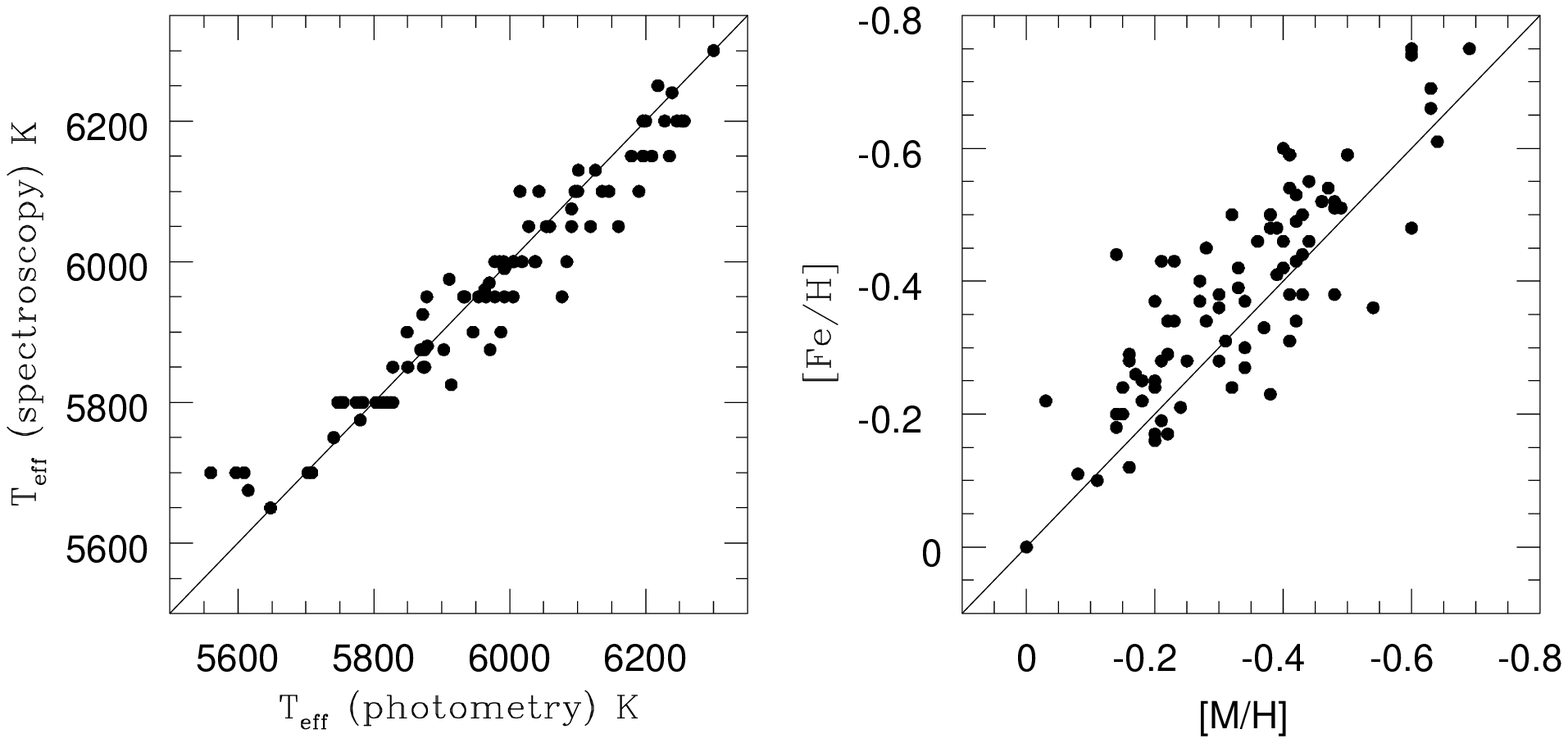}
\caption[] {Photometric temperature, $T_{\rm eff}$ and metallicity, [M/H] for a sample of 82 stars
are compared with our spectroscopically derived $T_{\rm {eff}}$ and [Fe/H] values
(see the
text for details).}
\end{figure*}

\clearpage
\begin{figure*}
\epsfxsize=18truecm
\epsffile{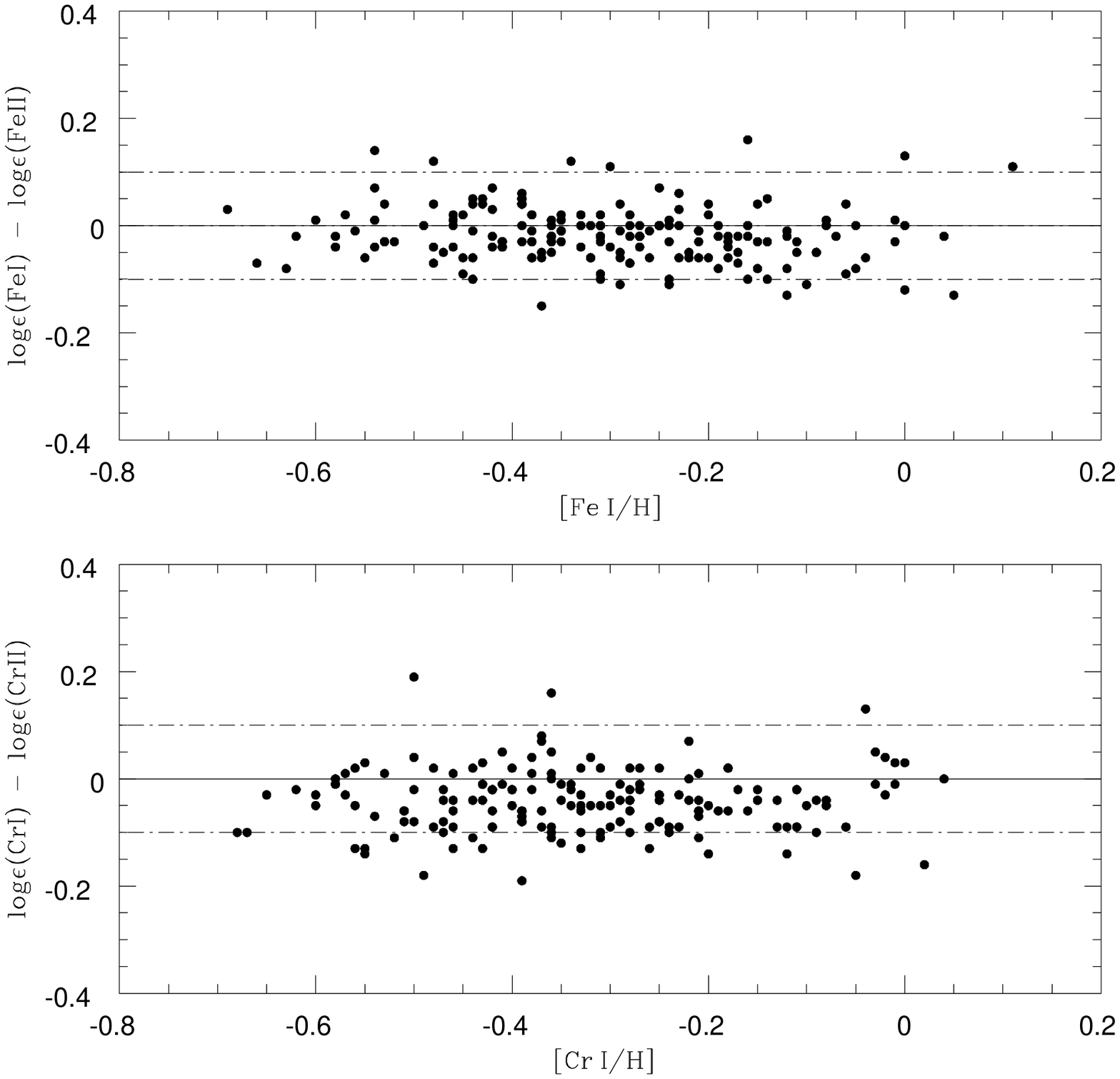}
\caption{The abundance differences from 
neutral and singly ionized lines of Fe and Cr are plotted against
the abundances from the neutral lines. The broken horizontal lines
represent the differences of $\pm$0.1 dex.}
\end{figure*}

\clearpage
\begin{figure*}
\centering
\includegraphics[width=8cm,angle=0]{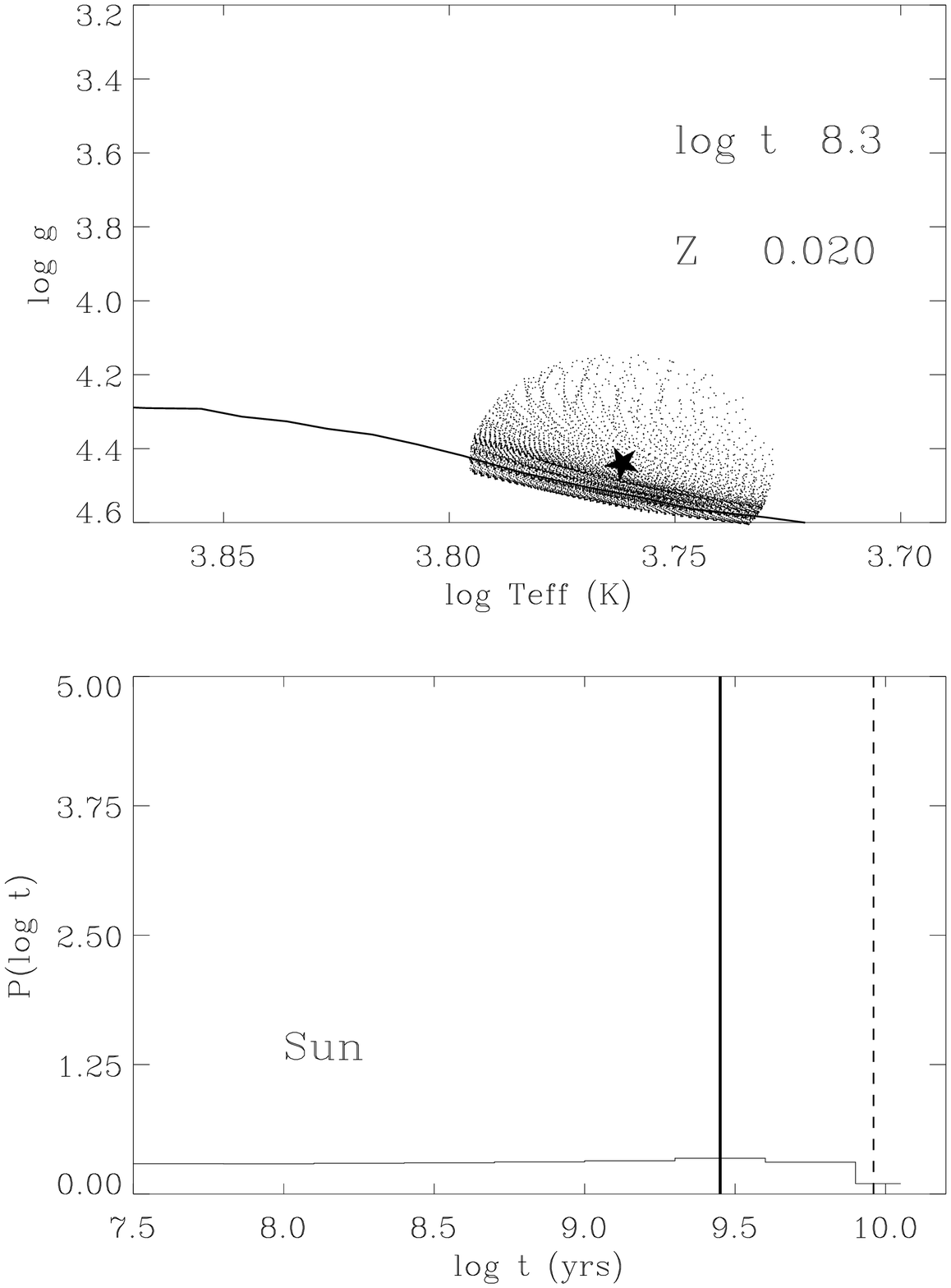}
\includegraphics[width=8cm,angle=0]{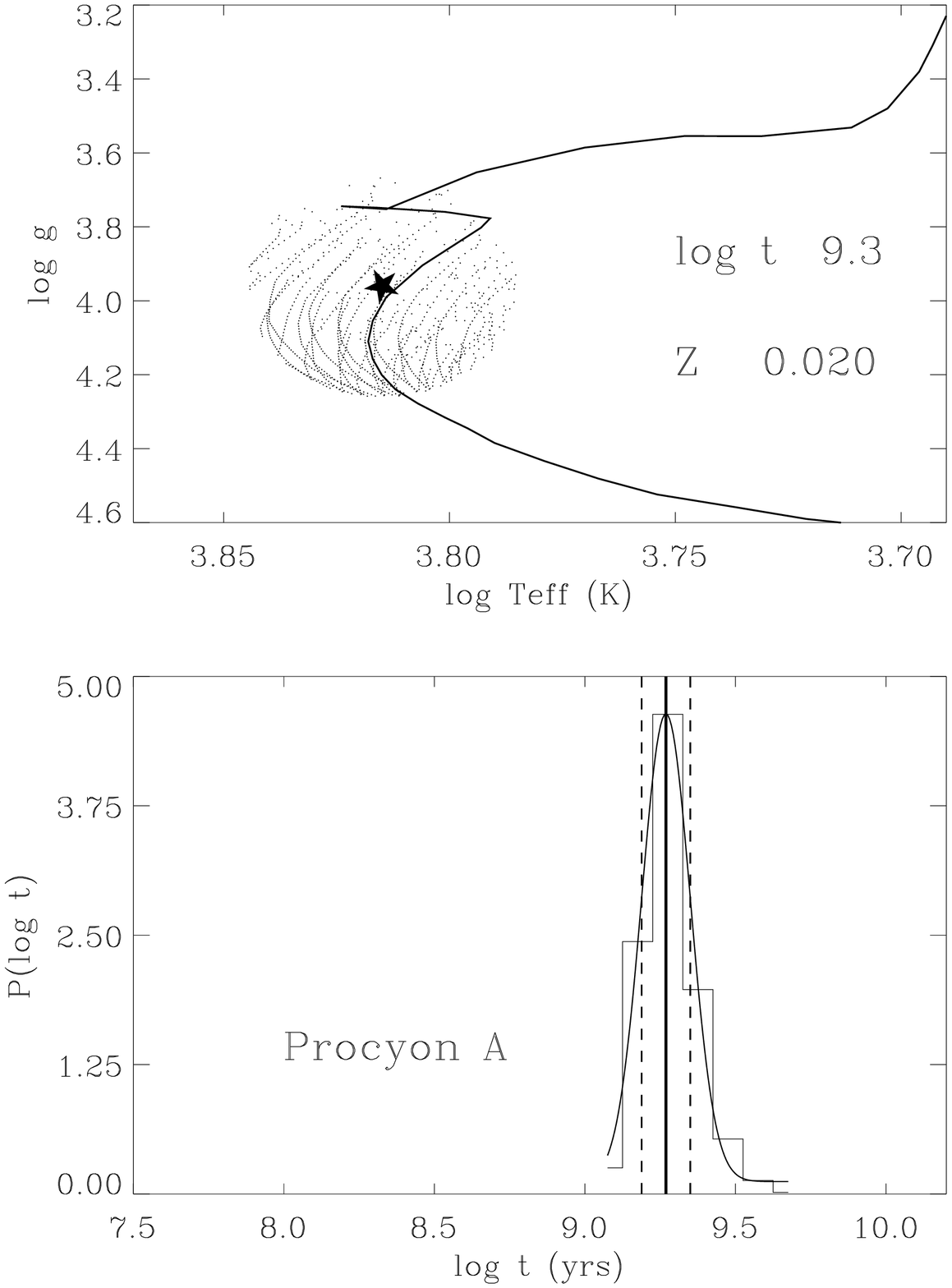}
\protect\caption[ ]{Upper panels: position of the star
in the $\log T_{\rm eff}- \log g$ plane. An isochrone with the indicated 
age and  metallicity is shown as a reference. The dots are grid points within 
the 3-$\sigma$ error bar ellipse. Lower panels: the probability density for 
the age is displayed.  The thick solid vertical line shows the best estimate
for the age and the broken vertical lines mark the 1-$\sigma$ limits.
}
\label{f1}
\end{figure*}

\clearpage
\begin{figure*}
\epsfxsize=18truecm
\epsffile{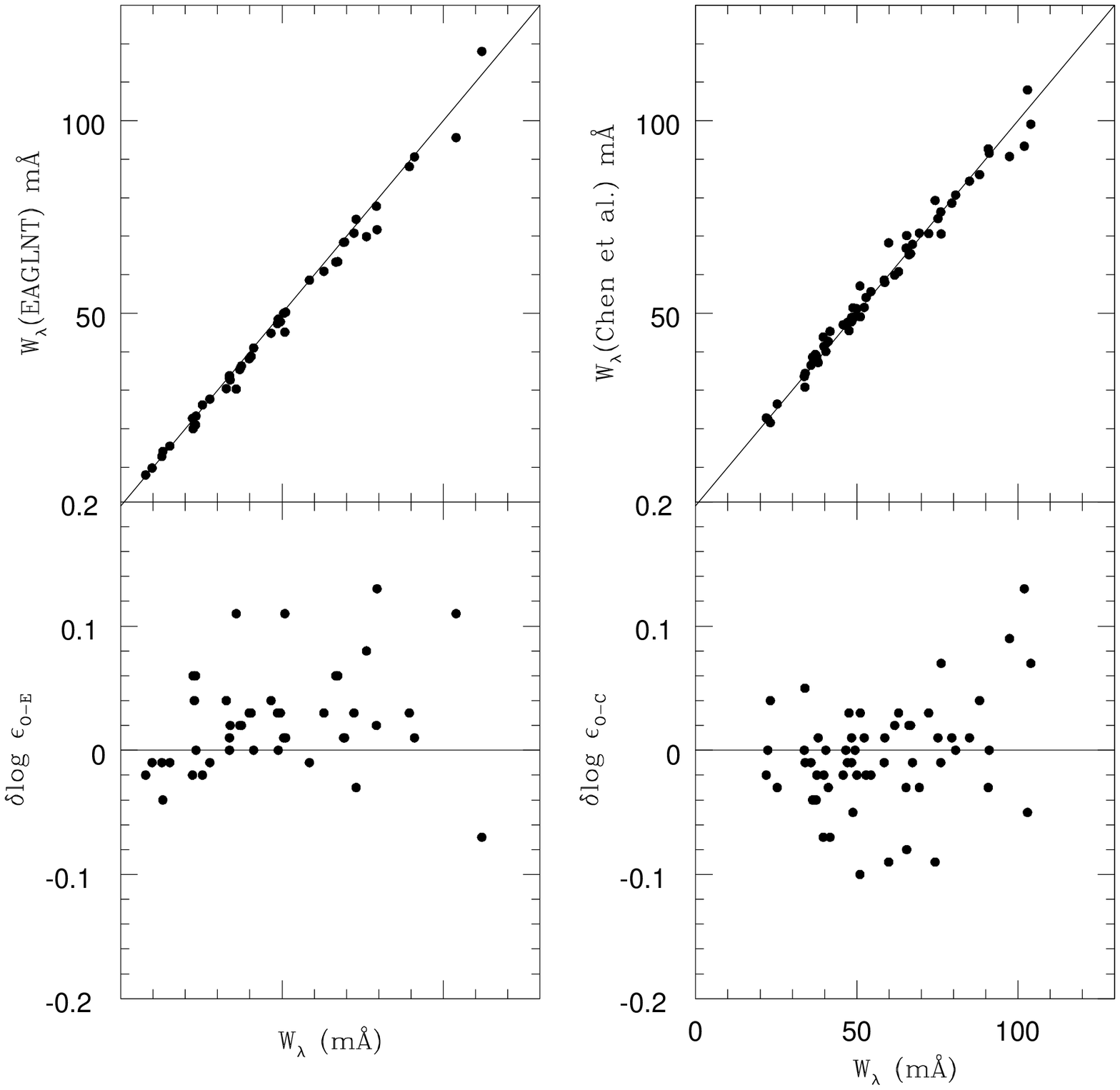}
\caption{The measured equivalent widths and the derived
abundances for the solar Iris spectrum are compared with
the solar measurements of EAGLNT (left panels) and Chen et al. (right panels) studies.}
\end{figure*}

\clearpage
\begin{figure*}
\epsfxsize=18truecm
\epsffile{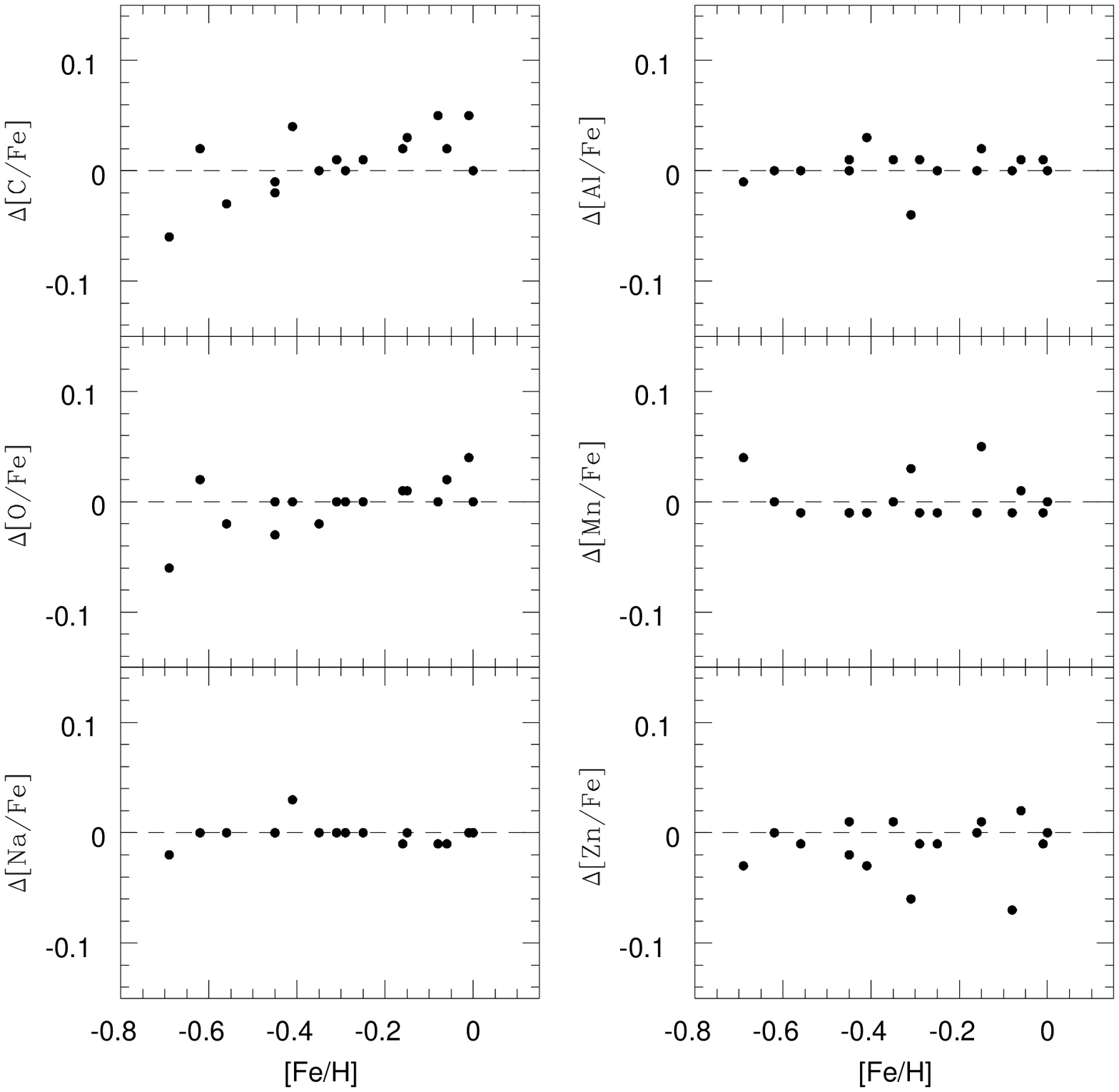}
\caption{Differences in [X/Fe] between OVER and NOVER models
for 16 stars. }
\end{figure*}

\clearpage
\begin{figure*}
\epsfxsize=18truecm
\epsffile{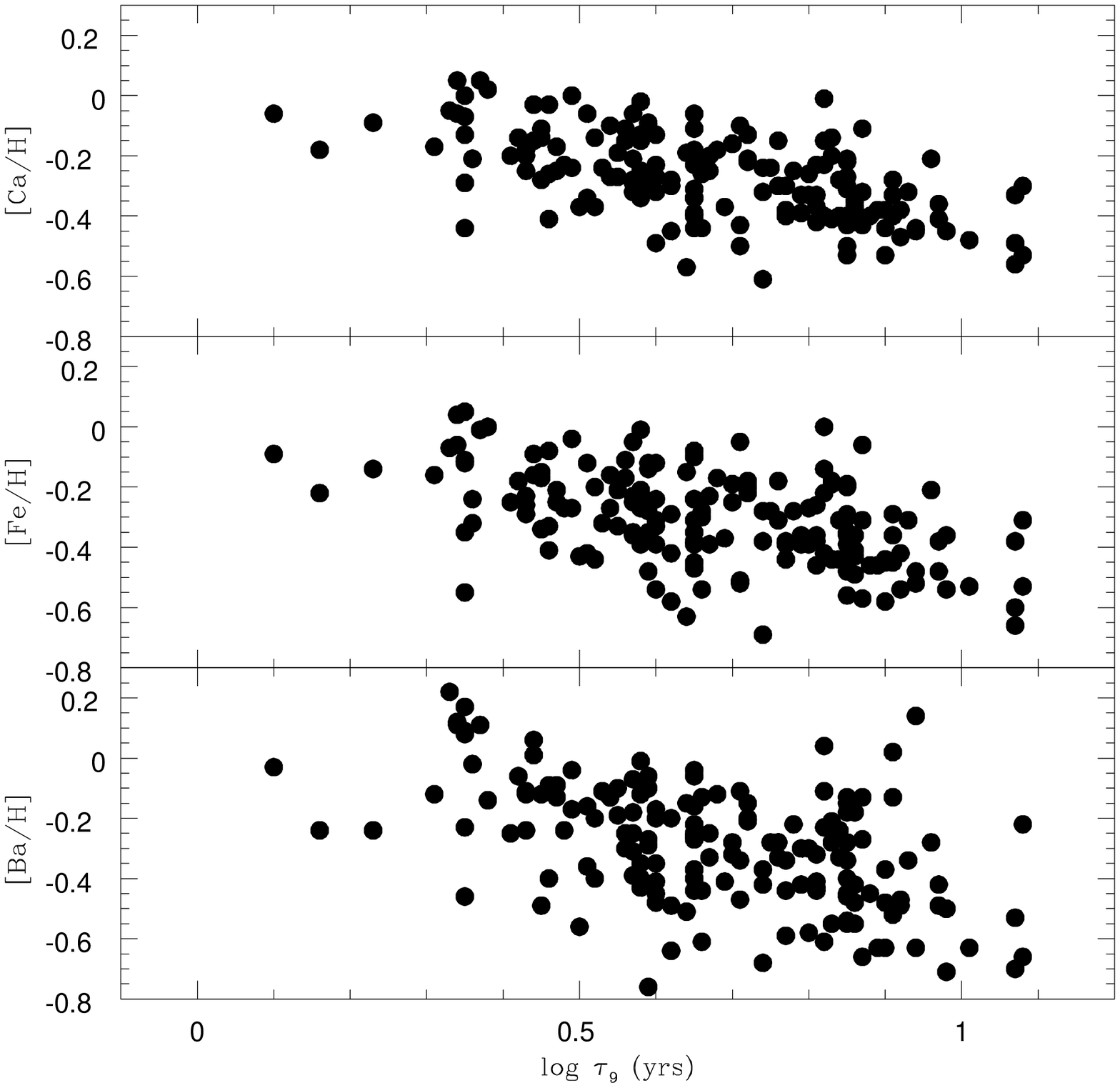}
\caption{ Relative abundances of Ca, Fe, and Ba
are shown against the ages of the stars.} 
\end{figure*}

\clearpage
\begin{figure*}
\epsfxsize=18truecm
\epsffile{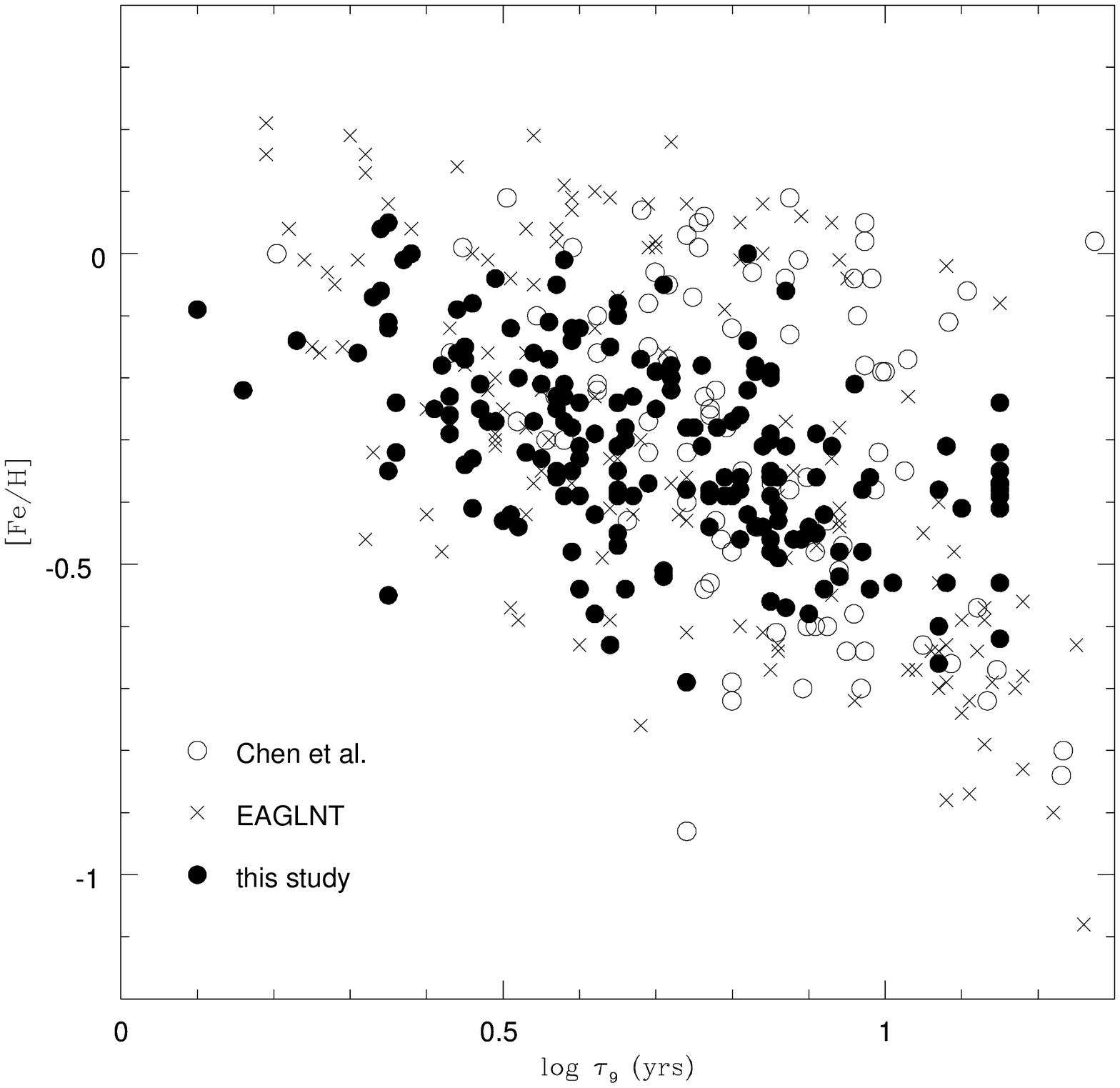}
\caption{ The Age-[Fe/H] relation derived in this study
is compared with the two previous studies of Chen et al. and EAGLNT.} 
\end{figure*}

\clearpage
\begin{figure*}
\caption{ Abundance ratios [X/Fe] from C to K, are plotted against the metallicity [Fe/H].
Note that oxygen abundances derived from permitted triplet lines at 7775~\AA\
are corrected using equation~3.}
\end{figure*}

\begin{figure*}
\caption{ Abundance ratios [X/Fe] from Ca to Zn,  are plotted against the metallicity [Fe/H].}
\end{figure*}

\begin{figure*}
\caption{ Abundance ratios, [X/Fe] from Sr to Eu,  are plotted against the metallicity, [Fe/H].}
\end{figure*}

\begin{figure*}
\caption{ Abundances of Mg and Si relative to Fe (top panels) are plotted against [Fe/H].
The solid line is the least squares fit to the data. The residuals around the mean fit
are shown in the middle panels. Bottom panels show histograms and the Gaussian fit
to the residuals.}
\end{figure*}

\clearpage
\begin{figure*}
\epsfxsize=18truecm
\epsffile{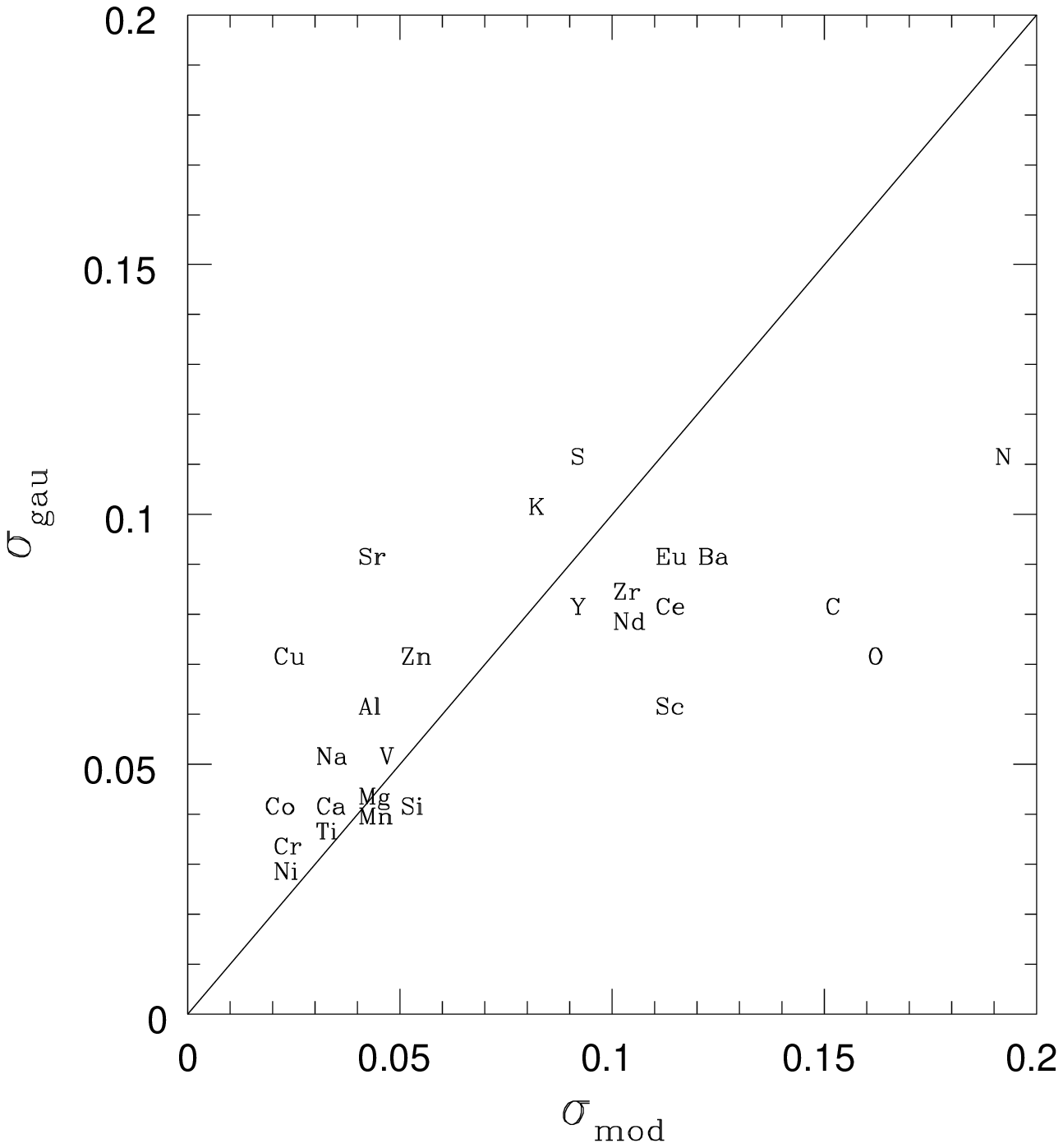}
\caption{ The errors in the abundance analysis as represented by $\sigma_{mod}$ are
compared with $\sigma_{gau}$, the dispersion of the Gaussian distribution of the abundance residuals.}
\end{figure*}

\clearpage
\begin{figure*}
\epsfxsize=18truecm
\epsffile{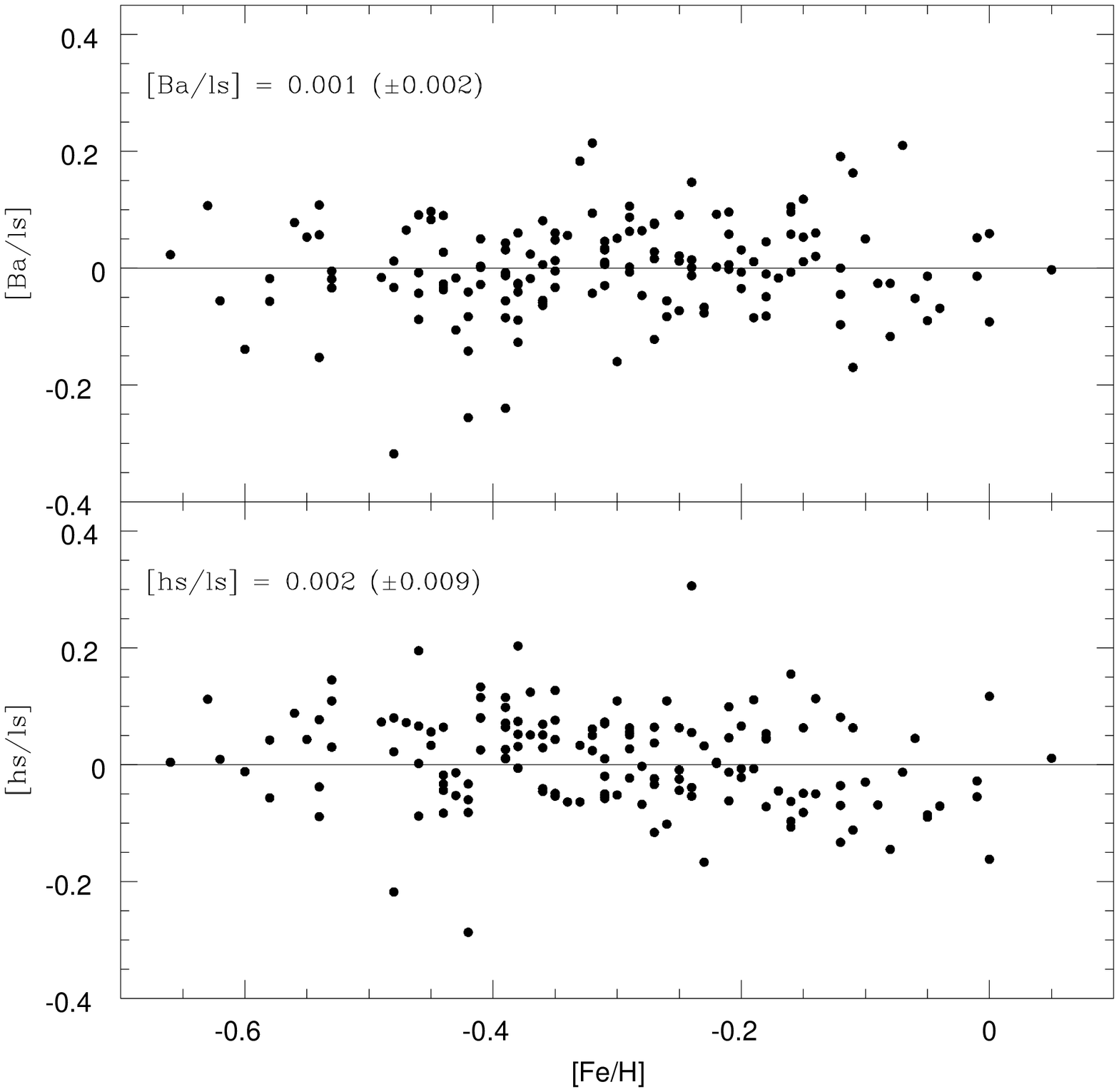}
\caption{ Abundance ratios of heavy (hs: Ba, Ce, and Nd) ) to light (ls: Sr and Y) $s$-process elements
for our sample of stars are shown against metallicity.}  
\end{figure*}

\clearpage
\begin{figure*}
\caption{Abundance ratios, [Mg/Fe] and [Ti/Fe] against [Fe/H].
Our data (filled circles)
is compared with earlier studies of Chen et al. (2000: open circles), EAGLNT (open squares),
Fuhrmann (1998: crosses), Fulbright (2000: stars), and Prochaska et al. (2000)(open triangles).
The broken horizontal lines are drawn for the mean abundances of Fuhrmann and Prochaska et al. thick
disc stars.}
\end{figure*}

\clearpage
\begin{figure*}
\caption{Plots of $V_{\rm LSR}$, $W_{\rm LSR}$, $R_{\rm m}$, and age
against [Fe/H]. Values determined in this study (filled circles) are
compared with earlier studies: Chen et al. (open circles), EAGLNT (open squares),
Fuhrmann (crosses), Fulbright (stars), and Prochaska et al. (open triangles). In all cases bigger
symbols represent stars with [Mg/Fe] $\geq$ 0.2.}
\end{figure*}

\clearpage
\begin{figure*}
\epsfxsize=18truecm
\epsffile{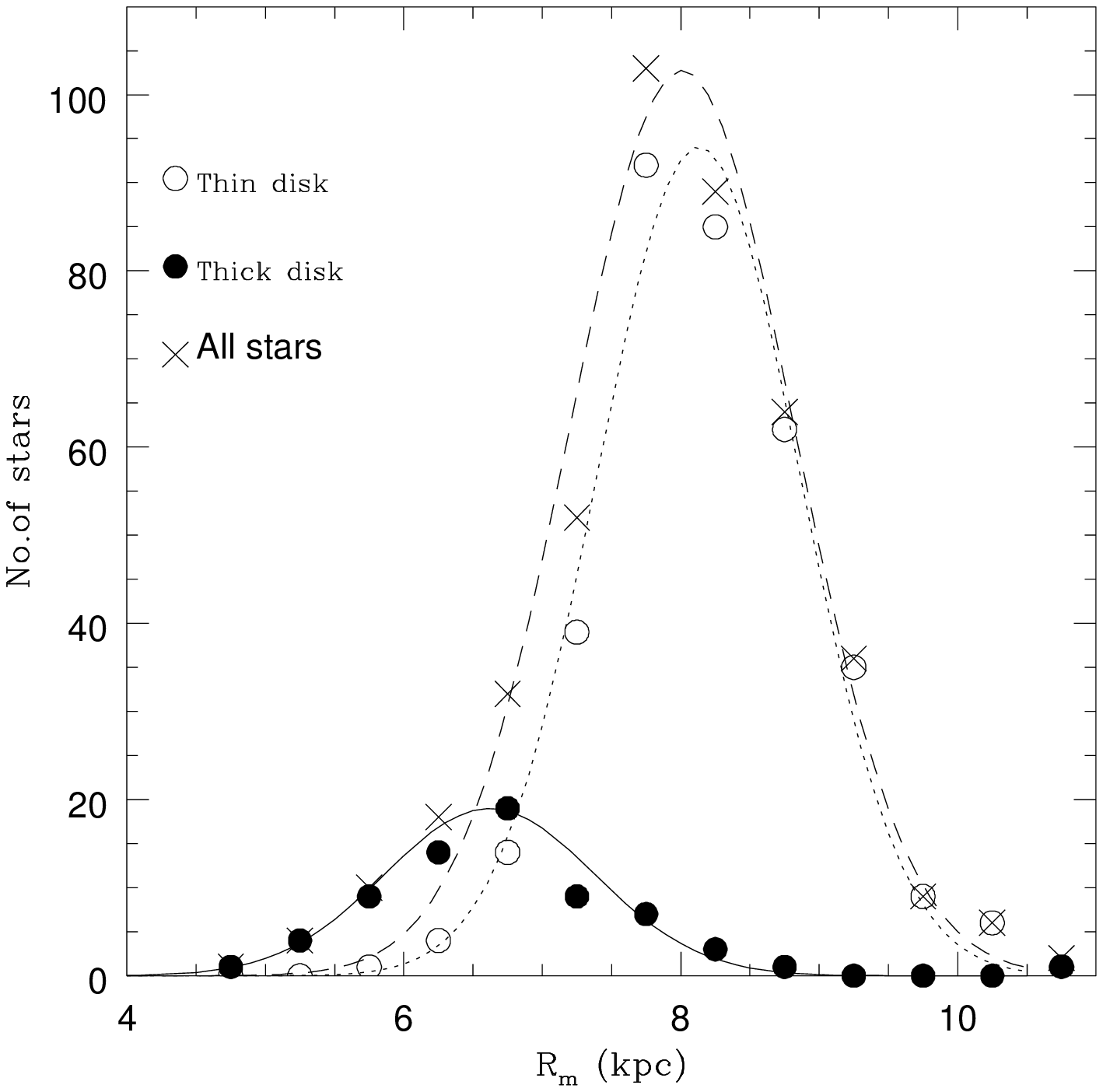}
\caption{ Distribution of around 500 disc stars in $R_{\rm m}$.
Sample is taken from three surveys: EAGLNT, Chen et al., Fulbright (2000), and
ours. The Gaussian fit to the entire sample (crosses) is
asymmetric (dashed line), however, fits are very symmetric for thin (dotted line),
and thick (solid line) disc populations.}
\end{figure*}

\clearpage
\begin{figure*}
\epsfxsize=18truecm
\epsffile{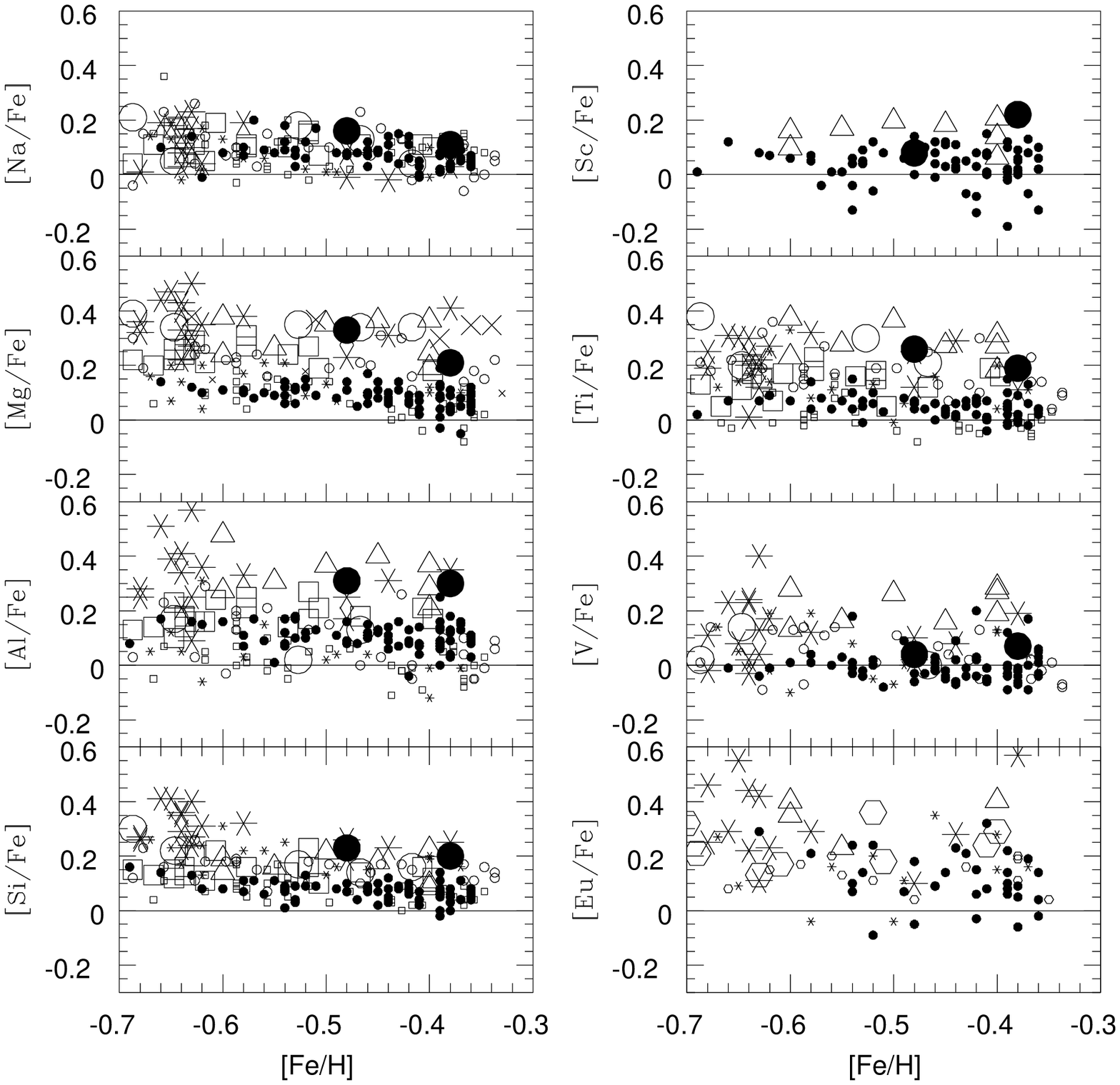}
\caption{Abundance ratios [X/Fe] of selected elements 
for stars in the current study (filled circles) in metallicity
range from $-$0.35 to $-$0.70, are compared with other studies: 
Chen et al. (2000: open circles), EAGLNT (open squares),
Fuhrmann (1998: crosses), Fulbright (2000: stars), Prochaska et al. (2000)(open triangles), 
and Woolf et al. (1995: hexagons). In all the cases bigger symbols represent stars
of thick disc.}
\end{figure*}

\clearpage
\begin{figure*}
\epsfxsize=18truecm
\epsffile{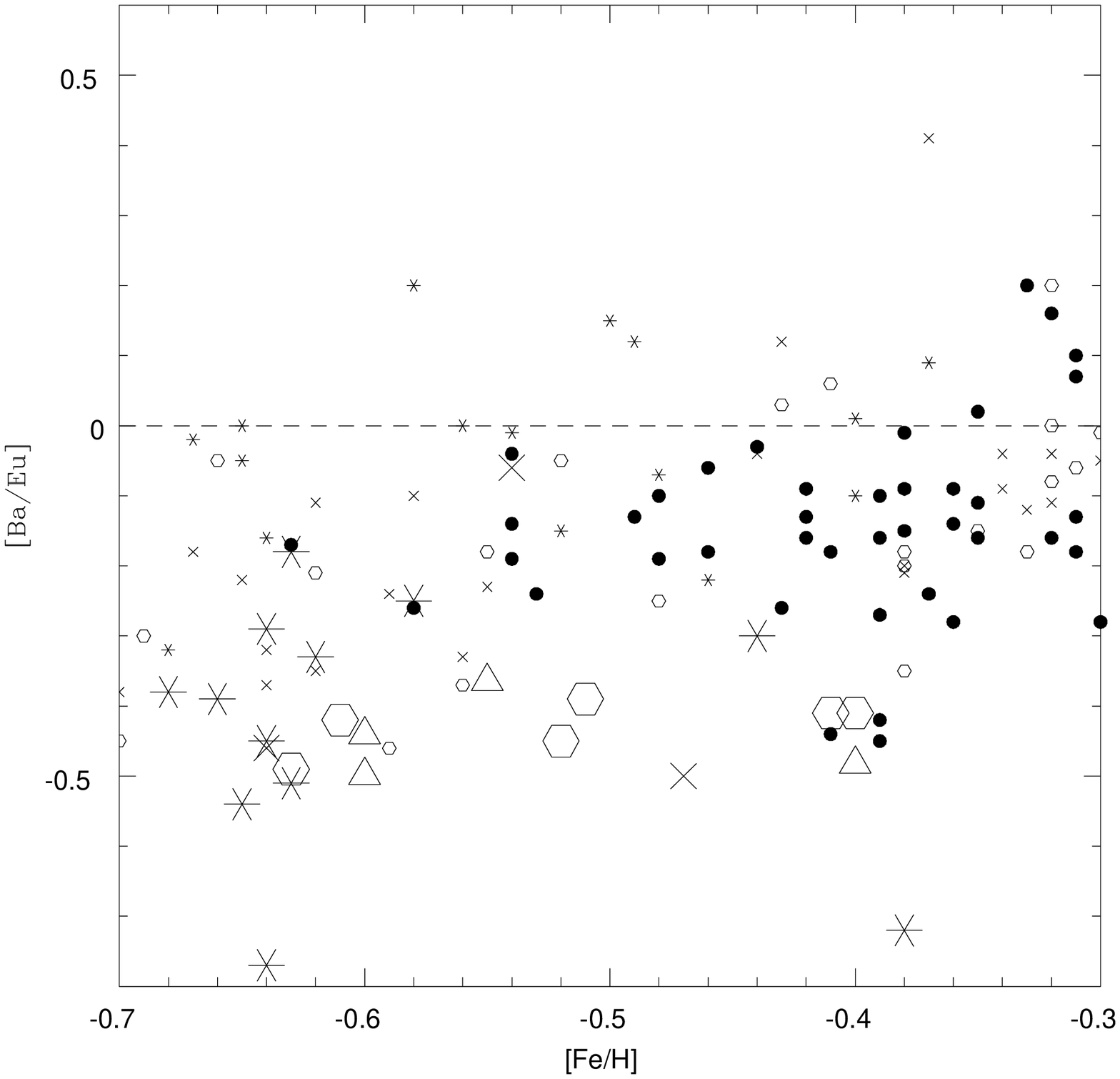}
\caption{ Abundance ratios [Ba/Eu] against [Fe/H] are shown. The symbols
represent same as in Figure\,18. For stars analysed by Woolf et al. (hexagons), and
Kock \& Edvardsson (2002: crosses) Ba abundances are taken from EAGLNT.}
\end{figure*}
\clearpage
\renewcommand{\thetable}{1}
\begin{table*}
\begin{center}
\caption{Atmospheric parameters and kinematic data for the programme stars.}\end{center}
\begin{tabular}{rccrrrrrrrr}
\hline \hline
Star   &  $T_{\rm eff}$ & log $g$ & [Fe/H] & $U_{\rm LSR}$ & $V_{\rm LSR}$ & $W_{\rm LSR}$ & $R_{\rm m}$ & 
$Z_{\rm max}$ & $e$ & log $\tau_{9}$ \\
HD     &   (K)    &(cm s$^{-2}$)&        & (km s$^{-1}$)& (km s$^{-1}$)& (km s$^{-1}$)& (kpc)  &(kpc) &  & (
yrs) \\
\hline
     70  &  5649.  &   4.37 &  -0.35 &  -16.7 &  -25.9 &  -29.5  &   7.59  &  0.241  &   0.13  &   ...\\
    101  &  5826.  &   4.36 &  -0.29 &  -55.7 &  -28.8 &   23.9  &   7.67  &  0.199  &   0.22  & 0.85\\
    153  &  5791.  &   3.80 &  -0.11 &   18.3 &  -43.0 &   9.9  &   7.11  & 0.075  &   0.21  &  0.56\\
    330  &  5775.  &   3.84 &  -0.27 &  -4.1  &  -50.2 &  -21.8  &   6.91  &  0.171  &   0.23  &  0.58\\
    912  &  6011.  &   3.82 &  -0.26 &   29.4 &  -5.5 &     6.8  &   8.26  & 0.053  &  0.10  &  0.43\\
   2663  &  5982.  &   4.33 &  -0.41 &   47.0 &  -11.9 &  -10.1  &   8.13  & 0.080  &   0.16  &  0.86\\
\end{tabular}
\end{table*}
\begin{table*}
\contcaption{}
\begin{tabular}{rccrrrrrrrr}
\hline \hline
Star   &  $T_{\rm eff}$ & log $g$ & [Fe/H] & $U_{\rm LSR}$ & $V_{\rm LSR}$ & $W_{\rm LSR}$ & $R_{\rm m}$ & 
$Z_{\rm max}$ & $e$ & log $\tau_{9}$ \\
HD     &   (K)    &(cm s$^{-2}$)&        & (km s$^{-1}$)& (km s$^{-1}$)& (km s$^{-1}$)& (kpc)  &(kpc) &  & (
yrs) \\
\hline
  86884  &  5873.  &   3.89 &  -0.28 &   34.20 &  -52.0 &  -22.4  &   6.93  &  0.179  &   0.26  &  0.59\\
  87838  &  6019.  &   4.29 &  -0.43 &   81.1 &  -9.1 &  -17.0  &   8.52  &  0.147  &   0.25  &  0.86\\
  88446  &  5875.  &   4.08 &  -0.35 &   33.0 &  -95.8 &    8.9  &   5.95  & 0.067  &   0.45  &  0.94\\
  89010  &  5611.  &   4.00 &  -0.06 &  -2.6  &   14.6 &  -36.9  &   8.97  &  0.329  &  0.05  &  0.87\\
  90878  &  6285.  &   3.85 &  -0.24 &   4.0  &  -8.3 &     2.9  &   8.11  & 0.021  &  0.05  &  0.36\\
  91638  &  6160.  &   4.29 &  -0.25 &  -24.6 &  -10.7 &   -9.7  &   8.07  & 0.074  &  0.09  &  0.57\\
\end{tabular}
\end{table*}
\begin{table*}
\contcaption{}
\begin{tabular}{rccrrrrrrrr}
\hline \hline
Star   &  $T_{\rm eff}$ & log $g$ & [Fe/H] & $U_{\rm LSR}$ & $V_{\rm LSR}$ & $W_{\rm LSR}$ & $R_{\rm m}$ & 
$Z_{\rm max}$ & $e$ & log $\tau_{9}$ \\
HD     &   (K)    &(cm s$^{-2}$)&        & (km s$^{-1}$)& (km s$^{-1}$)& (km s$^{-1}$)& (kpc)  &(kpc) &  & (
yrs) \\
\hline
 182758  &  5894.  &   4.13 &  -0.53 &  -26.8 &   5.3 &  -27.6   &   8.65  &  0.233  &  0.08  &   1.08\\
 186379  &  5806.  &   3.99 &  -0.39 &  -42.2 &  -21.5 &  -38.0  &   7.83  &  0.336  &   0.16  &  0.77\\
 186408  &  5670.  &   4.32 &   0.00 &  -28.0 &  -24.9 &   7.0   &   7.64  & 0.053  &   0.15  &  0.82\\
 190681  &  6105.  &   3.95 &  -0.08 &  -38.5 &  -21.2 &   3.7   &   7.79  & 0.028  &   0.16  &  0.46\\
 191649  &  5911.  &   4.23 &  -0.26 &   11.5 &  -1.3 &   3.2    &   8.36  & 0.024  &  0.04  &   0.81\\
 191672  &  5963.  &   4.10 &  -0.46 &   17.1 &  -18.0 &   2.6   &   7.82  & 0.019  &   0.10  &  0.89\\
\hline
\end{tabular}
\end{table*}

\clearpage
\renewcommand{\thetable}{2}
\begin{table*}
\begin{center}
\caption{Adopted line list and atomic data. $\Delta_{6}$ is the
enhancement factor with which classical van der Waals damping constant multiplied, and all other
columns are self explanatory.}\end{center}
\begin{tabular}{rccrrrr}
\hline \hline
Ion  &  Wavelength & LEP & log $gf$ & $\delta\Gamma_{6}^{5/2}$   & W$_{\lambda_{\odot}}$ & Ref. \\
& (\AA)    &       (eV)  &      &        &          (m\AA) & \\
\hline
C\,{\sc I} & 5052.150 &     7.68 & -1.30 &   9.9  &           37.8 & WFD\\
           & 5380.320 &     7.68 & -1.61 &   9.9  &           21.9 & WFD\\
           & 6587.620 &     8.54 & -1.00 &   9.9  &           14.0 & WFD\\
           & 7113.170 &     8.65 & -0.77 &   9.9  &           22.7 &WFD\\
           & 7115.170 &     8.64&  -0.93 &   9.9  &           26.7&WFD\\
           & 7116.960   &   8.65&  -0.91 &   9.9  &           18.5&WFD\\
\end{tabular}
\end{table*}

\begin{table*}
\contcaption{}
\begin{tabular}{rccrrrr}
\hline \hline
Ion  &  Wavelength & LEP & log $gf$ &   & W$_{\lambda_{\odot}}$ & Ref. \\
& (\AA)    &       (eV)  &      &        &          (m\AA) & \\
\hline
Ti\,{\sc i} & 5024.850 &       0.818 &  -0.56  &   9.9  &            73.2&OXF\\
 & 5113.450 &        1.443 &  -0.73  &   9.9  &            27.6 &OXF\\
 & 5219.710 &        0.021 &   -2.24  &   9.9  &            29.1&OXF\\
 & 5866.460 &        1.066 &  -0.76  &   9.9  &            48.7&OXF\\
 & 6091.180 &        2.267 &  -0.37  &   9.9  &            15.3&OXF\\
 & 6126.220 &        1.066 &   -1.37  &   9.9  &            23.2&AP\\
\end{tabular}
\end{table*}

\begin{table*}
\contcaption{}
\begin{tabular}{rccrrrr}
\hline \hline
Ion  &  Wavelength & LEP & log $gf$ &   &W$_{\lambda_{\odot}}$ & Ref. \\
& (\AA)    &       (eV)  &      &        &          (m\AA) & \\
\hline
 & 6581.214 &        1.480 &   -4.68 &    2.3   &           19.6&HAN\\
 & 6591.330 &        4.593 &   -1.95 &    2.3   &           10.7&AP\\
 & 6608.040 &        2.279 &   -3.91 &    2.3   &           18.8&AP\\
 & 6625.027 &        1.010 &   -5.34 &    2.3   &           16.6&OXF\\
 & 6699.142 &        4.590 &   -2.10 &    2.3   &           8.4&HAN\\
 & 6713.750 &        4.795 &   -1.39 &    2.3   &           21.7&AP\\
\end{tabular}
\end{table*}

\begin{table*}
\begin{minipage}{100mm}
\contcaption{}
\begin{tabular}{rccrrrr}
\hline \hline
Ion  &  Wavelength & LEP & log $gf$ &  & W$_{\lambda_{\odot}}$ & Ref. \\
& (\AA)    &       (eV)  &      &        &          (m\AA) & \\
\hline
Sr\,{\sc i}  &4607.340   &   0.000  &  0.28  &   9.9  &            46.9&MB\\
  &          &            &                 &        & \\
Y\,{\sc ii}  &4900.120     &    1.03 & -0.09 &    9.9      &        57.4&HANN\\
  &5087.430      &   1.084 &  -0.16 &    9.9   &           48.6&HANN\\
  &5200.420      &  0.992 &  -0.57 &    9.9   &           39.0&HANN\\
  &5402.780      &   1.839 &  -0.44 &    9.9   &           12.6&HANN\\
  &          &            &                 &        &  &\\
Zr\,{\sc ii}  & 5112.280 &    1.66  & -0.59  &   9.9   &           9.7&BTR\\
  &          &            &                 &        & \\
Ba\,{\sc ii}  & 5853.690  &    0.604 &  -0.91  &   9.9   &           66.7&DSV\\
  &6141.730   &     0.704  & -0.03   &  9.9    &          132.8&DSV\\
  &6496.910   &     0.604  & -0.41   &  9.9    &          112.2&DSV\\
  &          &            &                 &        && \\
Ce\,{\sc ii}  &4523.080   &     0.51  & 0.04   &  16.    &          14.4&AP\\
  &4628.160   &     0.52  &  0.23   &  16.    &          20.3&AP\\
  &4773.960   &     0.92  &  0.25   &  16.    &          10.5&AP\\
  &          &            &                 &        & &\\
Nd\,{\sc ii}  &5092.800   &     0.38  & -0.65  &  9.9    &          7.6&AP\\
  &5319.820   &     0.55  & -0.28   &  9.9    &          11.7&AP\\
  &          &            &                 &        & \\
Eu\,{\sc ii}  &6645.130 &      1.380  &  0.20   &  9.9    &          5.6&KUR \\
\hline \\
\end{tabular}
\raggedright
References for the $gf$ values:- AP: astrophysical values derived from inverting solar and stellar spectra;
BG: Bi\'{e}mont \& Godfroid (1980); 
BTR: ; 
CAR:Cardon et al. (1982); 
DSV: Davidson et al. (1992);
HAN: Group at Hannover (Bard et al. 1991; Bard \& Kock 1994); 
HANN: Hannaford et al. (1982); 
KUR: Kurucz (1998); 
LAM: Lambert (1978)
LD: Lawler \& Dakin (1989);
LL: Lambert \& Luck (1978);
LUCK: R.E. Luck (1997, private communication);
LW: Lambert \& Warner (1968); 
MB: Migdalek \& Baylis (1987);
MOR: Milford, O'Mara, \& Rose (1994) ; 
OXH: mean values of HANN and OXF;
OXF: Group at Oxford (Blackwell et al. 1995 references therein);
SR: Smith \& Raggett (1981);
WFD: Wiese, Fuhr \& Deters (1996); 
WHL: Whaling et al. (1985); 
WL: Wickliffe \& Lawler (1997); 
YA: Youssef \& Amer (1989)\\ 
\end{minipage}
\end{table*}

\clearpage
\renewcommand{\thetable}{5}
\begin{table*}
\caption{Abundance ratios [X/Fe] for elements from C to Ti for the programme stars}
\begin{tabular}{rccrrrrrrrrrrrr}
\hline \hline
HD & C & N  & O & [O] & Na & Mg & Al & Si & S & K & Ca & Sc & Ti \\
\hline
     70  &   0.21  &.... &    0.29 &-0.04&   0.02 &  -0.02  &  0.08  &  0.08  & 0.10 &    0.12  &  0.08  &  0.08  &  0.01\\
    101  &   0.19  &.... &    0.24 &0.09 &   0.01 &   0.01  &  0.04  &  0.05  &0.12 &    0.11  &  0.02  &  0.06  & -0.01\\
    153  &   0.28  &0.42 &    0.40 &0.01 &   0.11 &   0.06  &  0.15  &  0.11  &0.21 &    0.21  &  0.00  &  0.00  & -0.03\\
    330  &   0.33  &0.32 &    0.35 &0.12 &   0.05 &   0.07  &  0.09  &  0.07  &0.12 &    0.18  &  0.02  &  0.00  & -0.03\\
    912  &   0.15  &.... &    0.43 &.... &   0.08 &   0.09  &  0.08  &  0.08  & 0.16 &    0.34  &  0.06  & -0.03  &  0.04\\
   2663  &   0.20  &.... &    0.30 &.... &   0.05 &   0.07  &  0.04  &  0.04  &0.12 &    0.22  &  0.04  &  0.01  &  0.00\\
   3079  &   0.20  &0.43 &    0.27 &-0.07&   0.01 &  -0.03  &  0.04  &  0.03  &0.15 &    0.12  & -0.01  &  0.02  & -0.05\\
\end{tabular}
\end{table*}

\begin{table*}
\contcaption{}
\begin{tabular}{rccrrrrrrrrrrrr}
\hline \hline
HD & C & N  & O & [O] & Na & Mg & Al & Si & S & K & Ca & Sc & Ti \\
\hline
  90878  &   0.02  &.... &    0.39 &.... &   0.10 &   0.06  &  0.02  &  0.05  & 0.01 &    0.47  &  0.03  & -0.12  &  0.07\\
  91638  &   0.01  &.... &    0.22 &.... &   0.02 &   0.09  &  0.01  &  0.01  & -0.02 &    0.31  & -0.04  &  0.03  &  0.03\\
  94012  &   0.19  &.... &    0.33 &0.17 &   0.08 &   0.05  &  0.08  &  0.04  & 0.16 &    0.26  &  0.07  &  0.10  &  0.04\\
  94835  &   0.17  &0.32 &    0.19 &.... &   0.00 &   0.01  &  0.06  &  0.07  &0.16 &    0.02  & -0.05  &  0.12  & -0.05\\
  97037  &   0.07  &0.18 &    0.17 &.... &  -0.02 &   0.06  &  0.08  &  0.02  &0.10 &    0.11  & -0.01  &  0.06  &  0.00\\
  99126  &   0.12  &0.30 &    0.22 &.... &  -0.01 &   0.11  &  0.12  &  0.05  &0.05 &    ....  & -0.04  & -0.09  &  0.01\\
  99233  &   0.21  &.... &    0.38 &.... &  -0.01 &   0.10  &  0.15  &  0.08  &0.12 &    0.16  &  0.05  &  0.07  &  0.09\\
\end{tabular}
\end{table*}
\begin{table*}
\contcaption{}
\begin{tabular}{rccrrrrrrrrrrrr}
\hline \hline
HD & C & N  & O & [O] & Na & Mg & Al & Si & S & K & Ca & Sc & Ti \\
\hline
 191672  &   0.15  &.... &    0.30 & 0.18 &   0.07 &   0.09  &  0.11  &  0.08  &  0.09 &    0.34  &  0.08  &  0.04  &  0.06\\
 192145  &   0.14  &.... &    0.31 & .... &   0.13 &   0.09  &  0.09  &  0.07  &  0.16 &    0.33  &  0.03  &  0.02  &  0.00\\
 193664  &   0.13  &.... &    0.17 & 0.10 &  -0.05 &   0.00  &  0.01  &  0.02  &  0.18 &    0.06  & -0.02  &  0.02  & -0.03\\
 194497  &  -0.04  &-0.02&    0.22 & .... &   0.03 &  -0.01  & -0.01  &  0.03  & 0.00 &    0.46  &  0.04  &  0.05  &  0.07\\
 195200  &   0.16  &0.03 &    0.25 & .... &   0.01 &  -0.01  &  0.06  &  0.01  &  0.11 &    0.17  & -0.02  & -0.03  & -0.04\\
 198089  &   0.19  &.... &    0.30 & .... &   0.02 &   0.07  &  0.10  &  0.06  &  0.19 &    0.10  &  0.00  &  0.11  &  0.02\\
 198109  &   0.19  &.... &    0.29 & .... &   0.06 &   0.07  &  0.06  &  0.03  &  0.04 &    0.33  &  0.03  &  0.02  &  0.06\\
\hline
\end{tabular}
\end{table*}
\clearpage
\renewcommand{\thetable}{6}
\begin{table*}
\caption{Abundance ratios [X/Fe] for elements from V to Eu for the programme stars}
\begin{tabular}{rccrrrrrrrrrrrr}
\hline \hline
HD & V & Cr  & Mn & Co & Ni & Cu & Zn & Sr & Yr & Zr & Ba & Ce & Nd & Eu \\
\hline

     70  & -0.05  &  0.01 &  -0.08 &   0.02 &  -0.02 &  -0.03  &  0.11  & -0.12  & -0.05 &  ....   & -0.10  &  0.03  & ....  & 0.08\\
    101  & -0.06  & -0.02 &  -0.13 &   0.01 &  -0.02 &  -0.11  &  0.10  & -0.01  &  0.08 &   0.00  &  0.01  &  0.05  & 0.07  &....\\
    153  & -0.04  &  0.01 &  -0.07 &   0.00 &   0.01 &  -0.03  &  0.05  & -0.02  &  0.00 &  -0.03  & -0.19  & -0.16  &-0.10  & 0.01\\
    330  & -0.07  & -0.03 &  -0.15 &  -0.03 &  -0.03 &  -0.07  &  0.08  & -0.02  & -0.01 &   0.01  & -0.13  & -0.07  &-0.01  & 0.00\\
    912  & -0.05  & -0.02 &  -0.17 &  -0.08 &  -0.03 &  -0.14  &  0.00  &  0.10  &  0.07 &   0.04  &  0.02  &  0.00  &-0.06  & 0.01\\
   2663  & -0.05  & -0.05 &  -0.18 &  -0.04 &  -0.03 &  -0.13  &  0.07  &  0.02  &  0.01 &   ....  & -0.01  & -0.01  & 0.10  &....\\
   3079  & -0.03  & -0.06 &  -0.13 &  -0.04 &  -0.05 &  -0.12  &  0.06  & -0.02  & -0.01 &   0.02  & -0.09  & -0.05  & 0.03  & -0.06\\
\end{tabular}
\end{table*}
\begin{table*}
\contcaption{}
\begin{tabular}{rccrrrrrrrrrrrr}
\hline \hline
HD & V & Cr  & Mn & Co & Ni & Cu & Zn & Sr & Yr & Zr & Ba & Ce & Nd & Eu \\
\hline
  89010  & -0.07  & -0.03 &   -0.08 &   0.03 &   0.02 &   0.04  &  0.13  & -0.10  &  0.02 &   0.07  & -0.07  &  0.04  &  0.01  &   0.15\\
  90878  &  0.14  &  0.03 &   -0.13 &   0.04 &  -0.06 &  -0.18  & -0.01  & -0.52  &  .... &   ....  &  0.22  &  0.10  &  ....  &  ....\\
  91638  & -0.04  & -0.05 &   -0.12 &  -0.05 &  -0.03 &  -0.12  &  0.05  &  0.03  & -0.01 &  -0.12  &  0.00  & -0.03  &  ....  &   0.13\\
  94012  & -0.03  & -0.03 &   -0.13 &   0.01 &  -0.02 &  -0.09  &  0.09  & -0.06  &  0.01 &  -0.02  &  0.03  &  0.11  & -0.07  &....\\
  94835  & -0.08  & -0.03 &   -0.05 &   0.00 &   0.02 &   0.00  &  0.10  &  0.00  &  0.15 &   0.05  &  0.04  &  0.08  &  0.02  &....\\
  97037  & -0.04  & -0.02 &   -0.09 &  -0.02 &  -0.02 &  -0.03  &  0.09  & -0.14  & -0.02 &   ....  & -0.09  & -0.02  &  0.03  &  0.09\\
  99126  & -0.13  &  0.00 &   -0.13 &  -0.04 &   0.00 &  -0.13  &  0.01  &  ....  &  0.00 &  -0.08  &  0.00  &  0.01  &  ....  &   0.12\\
\end{tabular}
\end{table*}
\begin{table*}
\contcaption{}
\begin{tabular}{rccrrrrrrrrrrrr}
\hline \hline
HD & V & Cr  & Mn & Co & Ni & Cu & Zn & Sr & Yr & Zr & Ba & Ce & Nd & Eu \\
\hline
 191672  &  0.03  &  0.02 &   -0.16 &  -0.01 &  -0.01 &  -0.06  &  0.04  & -0.06  & -0.04 &  -0.14  & -0.17  & -0.17  &  ....  &....\\
 192145  & -0.02  & -0.01 &   -0.14 &   0.01 &   0.00 &  -0.04  &  0.08  &  0.02  & -0.03 &  -0.14  & -0.04  & -0.02  & -0.19  &....\\
 193664  & -0.10  & -0.02 &   -0.11 &  -0.08 &  -0.04 &  -0.11  &  0.04  &  0.01  &  0.10 &   0.08  &  0.06  &  0.16  &  0.16  &....\\
 194497  &  0.00  &  0.01 &   -0.18 &  -0.03 &  -0.05 &  -0.09  & -0.02  &  0.15  &  0.07 &  -0.12  &  0.17  &  0.12  &  0.12  &....\\
 195200  & -0.06  & -0.03 &   -0.12 &  -0.06 &  -0.05 &  -0.09  &  0.02  &  0.03  &  0.04 &  -0.01  &  0.05  &  0.01  & -0.02  &....\\
 198089  & -0.01  & -0.06 &   -0.13 &   0.00 &  -0.02 &  -0.03  &  0.10  & -0.15  & -0.03 &  -0.04  & -0.04  & -0.03  &  0.07  &   0.24\\
 198109  &  0.02  & -0.06 &   -0.17 &  -0.02 &  -0.05 &  -0.13  &  0.00  & -0.09  & -0.06 &  -0.06  & -0.11  & -0.03  &  0.01  &....\\
\hline
\end{tabular}
\end{table*}

\end{document}